\documentclass[11pt,a4paper]{article}
\pdfoutput=1
\usepackage[pdftex]{graphics}
\usepackage{jheppub}
\usepackage{amsmath,amssymb,amsfonts}
\usepackage{enumitem}
\usepackage{multirow}
\usepackage{array,booktabs}
\usepackage{tikz}
\usepackage{graphicx}
\usepackage[export]{adjustbox}
\usepackage{hyperref}

\newcommand{\be}{\begin{eqnarray}}
\newcommand{\ee}{\end{eqnarray}}
\newcommand{\nn}{\nonumber}
\newcommand{\bn}{\begin{enumerate}}
\newcommand{\en}{\end{enumerate}}

\newcommand{\eq}[1]{(\ref{#1})}

\def\identity{{\rlap{1} \hskip 1.6pt \hbox{1}}}
\def\iden{\identity}

\def\IN{\mathbb{N}}

\def\IR{\mathbb{R}}

\def\CA{{\cal A}}

\def\CD{{\cal D}}
\def\CE{{\cal E}}

\def\CH{{\cal H}}

\def\CO{{\cal O}}

\def\CW{{\cal W}}

\def\a{\alpha}
\def\b{\beta}
\def\g{\gamma}
\def\e{\epsilon}

\def\z{\zeta}
\def\th{\theta}

\def\l{\lambda}
\def\m{\mu}
\def\n{\nu}

\def\r{\rho}

\def\s{\sigma}

\def\t{\tau}

\def\G{\Gamma}
\def\D{\Delta}

\def\S{\Sigma}

\def\O{\Omega}

\def\half{\frac{1}{2}}

\def\imp{\Longrightarrow}

\def\del{\nabla}

\def\p{\partial}
\newcommand{\bra}[1]{\left\langle{#1}\right|}
\newcommand{\ket}[1]{\left|{#1}\right\rangle}

\def\identity{{\rlap{1} \hskip 1.6pt \hbox{1}}}

\newcommand{\bfig}{\begin{figure}}
\newcommand{\efig}{\end{figure}}

\def\la{{\langle}}
\def\ra{{\rangle}}
\def\abs#1{{\left| #1 \right|}}

\def\bl#1\el{\begin{align}#1\end{align}}

\newcommand{\fbra}[1]{ ({#1} |}
\newcommand{\fket}[1]{ | {#1} )}
\newcommand{\fbraket}[2]{ ( {#1} | {#2} )}

\begin{document}

\title{Explicit reconstruction of the entanglement wedge}

\author[a]{Jung-Wook Kim}

\affiliation[a]{Department of Physics and Astronomy, Seoul National University, Seoul 08826, Korea}

\emailAdd{jwkonline@snu.ac.kr}

\abstract{The problem of how the boundary encodes the bulk in AdS/CFT is still a subject of study today. One of the major issues that needs more elucidation is the problem of subregion duality; what information of the bulk a given boundary subregion encodes. Although the proof given by Dong, Harlow, and Wall\cite{Dong:2016eik} states that the entanglement wedge of the bulk should be encoded in boundary subregions, no explicit procedure for reconstructing the entanglement wedge was given so far. In this paper, mode sum approach to obtaining smearing functions for a single bulk scalar is generalised to include bulk reconstruction in the entanglement wedge of boundary subregions. It is generally expectated that solutions to the wave equation on a complicated coordinate patch are needed, but this hard problem has been transferred to a less hard but tractable problem of matrix inversion.}


\maketitle

\section{Introduction}
The AdS/CFT correspondence, first conjectured by Maldacena\cite{Maldacena:1997re}, states that the gravitational theory on a asymptotically anti-de Sitter spacetime is dual to the conformal field theory defined on its boundary. However, the exact nature of this duality is still not well understood.

One of the facets of this duality that remains elusive is emergence of a (quasi-)local point in the bulk when the dual boundary CFT is given. There is a vast amount of literature on the subject, some of which probe the bulk by examining the entanglement entropy and related quantities of the boundary.\footnote{Although emergence of the dual bulk theory is in principle independent of entanglement in the boundary theory, the insight gained by Van Raamsdonk\cite{VanRaamsdonk:2010pw} has tempted many researchers to equate the two.} For example, there are approaches utilising differential entropies\cite{Balasubramanian:2013lsa,Headrick:2014eia}, inverting the Ryu-Takayanagi formula to obtain bulk local data\cite{Faulkner:2013ica,Lin:2014hva,Lin:2015lfa}, and using entanglement entropy to derive linearised Einstein equations in the bulk\cite{Nozaki:2013vta,Lashkari:2013koa,Faulkner:2013ica,Swingle:2014uza}. A more complete list of references in related directions can be found in the thesis by Lin\cite{Lin:2015lfa}. Another line of research focuses on the causal structure of the bulk, using null geodesics\cite{Bousso:2012mh} or light-cones\cite{Engelhardt:2016wgb} that extend into the bulk. Other approaches utilise constraints imposed by CFT to probe bulk data\cite{Maldacena:2015iua,Czech:2016xec}. These approaches attempt to study reconstruction of the bulk metric\cite{Balasubramanian:2013lsa,Headrick:2014eia,Engelhardt:2016wgb}, the bulk matter fields\cite{Bousso:2012mh,Czech:2016xec}, or bulk dynamics\cite{Nozaki:2013vta,Lashkari:2013koa,Faulkner:2013ica,Swingle:2014uza,Maldacena:2015iua}.

In this paper, an answer is sought for the question `How is the bulk local (scalar) operator $\phi_P$ at bulk point $P$ represented by operators of the boundary?'. The more recent approaches outlined above are rather unwieldy for this purpose, so a different approach is pursued here. The approach of this paper is based on the construction of the dual bulk operator using smearing functions proposed by Hamilton, Kabat, Lifschytz, and Lowe(HKLL)\cite{Hamilton:2005ju,Hamilton:2006az}. Smearing functions are integral kernels integrated against the dual boundary operator, which depend on the patch used to reconstruct the bulk operator. For example, in global patch reconstruction the integration domain is taken to be boundary points that are spacelike separated from the bulk point being constructed, while in AdS-Rindler reconstruction the integration domain is the whole boundary covered by the coordinate patch.

On the other hand, there are attempts to understand the gravity dual of the CFT restricted to a subregion of the boundary\cite{Parikh:2012kg,Czech:2012be,Bousso:2012sj,Bousso:2012mh}, meaning that the integration domain of the smearing function is restricted to the boundary domain of dependence\footnote{Given a subregion $A$ of the boundary time slice $\S_b$, the \emph{boundary domain of dependence} $\CD[A]$ is the set of boundary points which any fully extended causal curve on the boundary that passes through the point must intersect $A$.} of the given subregion\cite{Almheiri:2014lwa}. Construction of bulk operators from a boundary subregion has an interesting property that while the bulk point $X$ can be constructed from subregion $A$ or subregion $B$, it may not be possible to construct the same bulk point from the intersection $A\cap B$. This observation had led to the conjecture that the AdS/CFT correspondence has a quantum error correcting code-like structure, which was proposed by Almheiri, Dong, and Harlow(ADH)\cite{Almheiri:2014lwa}. These discussions were mostly based on AdS/CFT correspondence for a \emph{connected} subregion of the boundary.

The most straightforward way of constructing the bulk from \emph{disconnected} boundary subregions is to decompose the disconnected set into connected parts and construct the bulk from each connected parts. This can be called the \emph{causal wedge reconstruction} as all bulk points lie in the causal wedge\footnote{The \emph{causal wedge} $\CW_C[A]$ of the given subregion $A$ is the set of bulk points that can receive and transmit light signals to its boundary domain of dependence $\CD[A]$\cite{Almheiri:2014lwa}.} of the boundary subregion\cite{Bousso:2012sj,Bousso:2012mh,Czech:2012be,Leichenauer:2013kaa,Engelhardt:2016wgb}. Nevertheless, a recent conjecture in the literature states that the bulk dual larger than the causal wedge can be constructed from the boundary subregion\cite{Czech:2012bh,Dong:2016eik,Bao:2016skw}, and this conjecture has been claimed prooved by Dong, Harlow, and Wall(DHW)\cite{Dong:2016eik}. How this construction can be explicitly done in a way similar to the HKLL procedure, unfortunately, is not properly understood yet\cite{Dong:2016eik,Jafferis:2015del}. This is the problem that this paper attempts to solve for the case of a scalar field of the bulk.

The goal of this paper is to suggest a procedure that explicitly constructs bulk operators from operators defined on boundary subregions, provided the global patch dictionary of the duality is already given in terms of mode operators and mode functions. The key tactic is just a small variant of mode sum approach, a widely used method to construct the bulk from the boundary\cite{Hamilton:2005ju,Hamilton:2006az,Banks:1998dd,Bena:1999jv,Bousso:2012mh}. The key feature that distinguishes this approach from others found in the literature is that it is based on appropriate rearrangement of mode functions, rather than obtaining a new mode function in a new coordinate patch and using it to construct smearing functions\cite{Hamilton:2006az,Bousso:2012mh,Almheiri:2014lwa,Leichenauer:2013kaa}. While new mode functions obtained by solving the wave equation on a new coordinate patch are orthogonal to each other, this is not the case for the mode functions obtained from the procedure outlined in this paper. The problem simplifies to the problem of obtaining coefficients in a non-orthogonal basis, where basis vectors are simply the mode functions at the boundary.

Based on the observation that a free scalar field in pure AdS can be identified as a generalised free field(GFF) of the boundary, the dictionary between mode operators and mode functions of the bulk and boundary is constructed for use as a working example. Using the dictionary and Hilbert space decomposition, smearing functions that reconstruct the bulk from given boundary subregions is obtained. This means that the proposed construction for the example only concerns linearised perturbations around pure AdS. Although this may sound uninteresting as a plethora of literature dealing with linearised perturbations in diverse directions exists\cite{Lin:2014hva,Nozaki:2013vta,Lashkari:2013koa,Faulkner:2013ica,Swingle:2014uza,Czech:2016xec}, this construction is a generalisation of the starting point for other interesting extensions found in the literature. Examples include incorporation of $1/N$ corrections and interactions\cite{Kabat:2011rz,Kabat:2015swa,Kabat:2016zzr}, and construction on a black hole background geometry\cite{Hamilton:2006fh,Papadodimas:2012aq,Leichenauer:2013kaa,Rey:2014dpa}. Thus, though the proposed construction is based on the mode sum approach which is rather old compared to more recent approaches, this is an approach that is still worthy of pursuit in that well-known interesting extensions exist.

All fields in this paper will be considered as operator \emph{distributions}, i.e. fields will acquire meaning as operators only when integrated against a suitable test function. For example, the local bulk operator $\phi_P$ mentioned earlier in this introduction is defined as $\phi_P = \int \eta_P(x) \phi(x) dx$, where $\eta_P(x)$ is a $\cal{C}^\infty$ function having compact support on a neighbourhood of bulk point $P$. This allows manipulations not allowed in functions and circumvention of miscellaneous complications. For those who are not familiar with distribution theory, \cite{Richards:distr} is a good introduction to the subject. \cite{Streater:pct} also has a short introduction to the theory of distributions.

This paper is organised as follows. Section \ref{sec:bulk-boundary} attempts to establish a concrete foundation for the bulk-boundary dictionary for perturbations of pure AdS, which is used as a working example for the arguments of this manuscript. Section \ref{sec:bulkrecon} reviews the bulk reconstruction algorithm of HKLL\cite{Hamilton:2006az} and generalises the procedure to cases which HKLL fails to give a reconstruction. Section \ref{sec:bulksubregion} provides an explicit decomposition of the bulk Hilbert space into bulk subregion Hilbert spaces in the non-interacting limit, which is needed in the bulk reconstruction of the entanglement wedge. Section \ref{sec:entwedgerecon} reviews the claim of DHW\cite{Dong:2016eik} that bulk reconstruction of entanglement wedge is possible and describes how the main goal of this paper can be achieved. The paper ends with section \ref{sec:discussion} which mulls over subtleties and future prospects of the construction.

\section{Non-interacting bulk limit} \label{sec:bulk-boundary}
This section is intended to provide a concrete working example for the construction proposed in this paper. In the non-interacting limit, also known as infinite $N$ limit, bulk fields behave as free fields on $AdS$, while boundary fields behave as special limits of CFT operators called \emph{generalised free fields}. It will be argued that an one-to-one correspondence between these fields can be constructed for perturbations of the vacuum, giving the dictionary for constructing the bulk from arbitrary boundary subregions. This section can be skipped for readers willing to accept the dictionary \eq{eq:dictionary1}, \eq{eq:dictionary2}, and \eq{eq:dictionary3}.

\subsection{Free fields on AdS}
Much of this subsection resembles the materials of chapter 4 and 5 of \cite{Kaplan:lec}. Consider a free scalar field in AdS$_{d+1}$ of AdS radius 1.
\bl
ds^2 &=\frac{1}{\cos^2 (\r)}\left( -dt^2 + d\r^2 + \sin^2(\r) d\O_{d-1}^2 \right) \label{eq:adsmetric}
\\ S &=  \int \sqrt\abs{g} d^{d+1}x \left( -\half g^{\m\n}\del_\m\phi \del_\n\phi -\half  m^2 \phi^2 \right)
\el
The classical equation of motion for this action is the Klein-Gordon equation. The following mode decompositions provide an orthogonal set of solutions that correspond to normalisable modes\footnote{Normalisable modes correspond to Dirichlet boundary conditions for the spatial coordinate $\r$. Since normalisable modes relate the bulk operator to the boundary operator and non-normalisable modes act as sources, normalisable modes are needed.}\footnote{For tachyonic fields above the Breitenlohner-Freedman bound, non-normalisable modes could be needed. Such cases are not considered in this paper.} which do not diverge at $\r=0$ \cite{Hamilton:2006az,Bousso:2012mh,Kaplan:lec,Nastase:adscft}.\footnote{Hypergeometric functions are better for examining the wave equation, while Jacobi polynomials are more suited for calculations because of their orthogonality. Their orthogonality relations can be found in \cite{Abramowitz:hmf} or \cite{NIST:DLMF}.}
\bl
\phi(t, \r, \O) &= \sum_{n,l,m} a_{nlm} ~e^{-i(\D+2n+l)t} Y_{lm}(\O) \sin^l {\r} \cos^\D{\r} ~{}_2F_1(-n,n+\D+l;l+\frac{d}{2};\sin^2{\r})
\\ &= \sum_{n,l,m} \tilde{a}_{nlm} ~e^{-i(\D+2n+l)t} Y_{lm}(\O) \sin^l {\r} \cos^\D{\r} ~ P^{(l+d/2-1, \D-d/2)}_n(\cos{(2\r)})
\el
$n \in \IN = \{ 0,1,2,\cdots \}$. $l$ is the total angular momentum of the corresponding mode and $m$ collectively refers to quantum numbers that are needed to specify a mode. $\D$ is the conformal dimension of the dual CFT$_d$ operator and is given by $\D=d/2+\sqrt{m^2+d^2/4}$, the greater solution to $m^2=\D(\D-d)$. Quantisation leads to the following mode decomposition.
\bl
\phi(t, \r, \O) &= \sum_{n,l,m} a_{nlm} f_{nlm}(t, \r, \O) + a_{nlm}^\dagger f_{nlm}^\ast (t, \r, \O) \label{eq:bulkmodeexp1}
\\ [a_{nlm},a_{n'l'm'}^\dagger] &= \delta_{nn'}\delta_{ll'}\delta_{mm'}
\\ [a_{nlm},a_{n'l'm'}] & =[a_{nlm}^\dagger,a_{n'l'm'}^\dagger]=0
\el
Each $a_{nlm}^\dagger$ raises the energy of the state by $\D+2n+l$. Note that the number of distinct states with energy $\D+k$, $k \in \IN$, is given by $(k+n-1)!/[(n-1)! k!]$.

The inner product between mode functions\cite{Kaplan:lec} is defined as follows, and is used to determine the normalisation of the mode functions.
\bl
\fbraket{ f }{ g }_\S &\equiv i \int_{\S}d^{d}y \sqrt{\abs{h}} n^\m \left( f^\ast \del_\m g- g\del_\m f^\ast \right) \label{eq:fbraketdef}
\\ \fbraket{ f_{nlm}}{ f_{n'l'm'} }_\S &= - \fbraket{ f_{nlm}^\ast}{ f_{n'l'm'}^\ast }_\S = \delta_{nn'}\delta_{ll'}\delta_{mm'}
\\ \fbraket{ f_{nlm} }{ f_{n'l'm'}^\ast }_\S &= 0
\el
This is a slightly unconventional notation employed to exploit the full power of Dirac's bra-ket notation. $\S$ denotes the spacelike surface on which the inner product is evaluated, $y^a$ is the coordinate system that covers the submanifold $\S$, $n^\m$ is the unit future-directed normal($n^\m n_\m=-1$) to $\S$, and $h$ is the determinant of the induced metric $h_{ab}=\frac{\p x^\m}{\p y^a}\frac{\p x^\n}{\p y^b}g_{\m\n}$. Note that the inner product satisfies the following relations.
\bl
\left[ \fbraket{ f }{ g }_\S \right]^\ast = \fbraket{ g }{ f }_\S  = - \fbraket{ f^\ast }{ g^\ast }_\S \label{eq:innerrel}
\el
For reference, an explicit expression for $f_{nlm}(t, \r, \th)$ is given.
\bl
f_{nlm}(t, \r, \O) &= \frac{1}{N_{nlm}}e^{-i(\D+2n+l)t} Y_{lm}(\O) \sin^l {\r} \cos^\D{\r} ~ P^{(l+d/2-1, \D-d/2)}_n(\cos{(2\r)}) \label{eq:bulkmodeexp2}
\\ N_{nlm} & = \sqrt{\frac{\G(n+l+d/2)\G(n+\D-d/2+1)}{n! \G(n+l+\D)}}
\el

\subsection{Generalised free theory} \label{sec:gffmodes}
Generalised free fields(GFF) are defined as operators whose correlators factorise into a sum of products of two-point functions.
\bl
\la\CO(x_1)\CO(x_2)\ra & \propto \frac{1}{(x_1-x_2)^{2\D}}
\\ \la\CO(x_1)\CO(x_2)\CO(x_3)\cdots\CO(x_{2n})\ra &= \la\CO(x_1)\CO(x_2)\ra \cdots \la\CO(x_{2n-1})\CO(x_{2n})\ra + \text{permutations}
\el
This is a sort of an infinite $N$ limit of a CFT\cite{ElShowk:2011ag}. Generalised free theory(GFT) is a theory purely consisting of GFFs. The reason it is called generalised \emph{free} is because its Hilbert space has a Fock space like structure, which is the Hilbert space of a free theory. A GFT is \emph{not} a CFT because it does not have a stress tensor in its spectrum\footnote{This is because existence of $T_{\m\n}$ spoils the factorisation of correlators. Nevertheless, a GFT can be considered as an effective description of operators with small scaling dimensions in a CFT in the infinite $c$ limit\cite{ElShowk:2011ag}.}, but shares many properties that a CFT has\cite{ElShowk:2011ag,Kaplan:lec}. In a Euclidean theory with radial quantisation, for example, an operator can be inserted at the origin to produce a state with the same conformal dimension; operator-state correspondence exists just as in a CFT.
\bl
\CO(0) \ket{0} &\to \ket{\CO} \label{eq:operatorstate1}
\\ (P^2)^n (P_\m)^l \CO(0) \ket{0} &\to \ket{(P^2)^n (P_\m)^l \CO} \label{eq:operatorstate2}
\el
The expression $(P_\m)^l$ denotes a traceless symmetric combination of $l$ $P_\m$ operators that corresponds to some generalised spherical harmonic of angular momentum $l$. The state $\ket{\CO}$ has conformal dimension $\D$, while the state $\ket{(P^2)^n (P_\m)^l \CO}$ has conformal dimension $\D+2n+l$ and spin $l$. Note that this spectrum matches exactly with the spectrum formed by a single mode excitation in AdS free scalar theory. Imposing bootstrap constraints requires product operators such as $:\CO\CO:$, which are devoid of anomalous dimensions and can be matched to multiple mode excitations in AdS free scalar theory, to exist in a GFT\cite{ElShowk:2011ag}.

To compare a GFF of a GFT to a free field of a free theory on AdS, it is useful to have a mode expansion of a GFF. The operator-state correspondence relations \eq{eq:operatorstate1} and \eq{eq:operatorstate2} can be used to expand a GFF in terms of mode operators.
\bl
\CO(x) &= \sum \CO_{i_1, \cdots, i_d}^\dagger \frac{1}{M_{i_1, \cdots, i_d}}\left[ \left( x^1 \right)^{i_1} \times \cdots \times \left( x^d \right)^{i_d} \right] + \text{h.c.}
\el
The mode operator $\CO_{i_1, \cdots, i_d}^\dagger$ acts on the vacuum to create a state corresponding to $\ket{(P_1)^{i_1}\cdots(P_d)^{i_d}\CO}$, while Hermitian conjugate part annihilates the vacuum. The state created by the product of operators $:\CO_1 \CO_2:$, where $\CO_i$s represent primary and descendent states collectively, can be obtained from the vacuum by acting mode operators successively.
\bl
\ket{\CO_1} = \CO_1^\dagger \ket{0}, \ket{\CO_2} = \CO_2^\dagger \ket{0} \imp \ket{: \CO_1 \CO_2 :} = \CO_1^\dagger \CO_2^\dagger \ket{0}
\el
It can be shown that the only algebra that can be satisfied by mode operators is the following algebra up to normalisation, which is absorbed by normalisation factors $M_{i_1, \cdots, i_d}$\footnote{The first line is established by examining orthogonality of states. The second line follows from commutativity of scalars. Because radial ordering is implicitly assumed in radial quantisation, scalars always commute.}.
\bl
[\CO_{i_1, \cdots, i_d}, \CO_{i_1', \cdots, i_d'}^\dagger] &= \delta_{i_1 i_1'} \cdots \delta_{i_d i_d'} \label{eq:boundmodealgebra1}
\\ [\CO_{i_1, \cdots, i_d}, \CO_{i_1', \cdots, i_d'}] &= [\CO_{i_1, \cdots, i_d}^\dagger, \CO_{i_1', \cdots, i_d'}^\dagger] = 0 \label{eq:boundmodealgebra2}
\el
Using polar coordinates and rearranging the modes, this mode expansion can be tidied up into a form involving spherical harmonics\footnote{The reason for taking the complex conjugate of spherical harmonics will become clear soon.}.
\bl
\CO(r, \O) &= \sum \CO_{nlm}^\dagger \frac{ r^{2n+l} ~Y_{lm}^\ast (\O) }{M_{nlm}} + \text{h.c.}
\el
The algebra of mode operators is not very different from \eq{eq:boundmodealgebra1} and \eq{eq:boundmodealgebra2}. Up to now, the GFT being considered is defined on the manifold $\IR^d$. To set up a GFT on the manifold $\IR \times S^{d-1}$, the exponential map $r=e^\t$ and Weyl rescaling $g \to g' = e^{-2\t}g$ is introduced. This means the GFF $\CO$ undergoes a conformal rescaling $\CO \to \CO' = (e^\t)^\D \CO$ as well\footnote{Introduction of conformal rescaling induced by Weyl rescaling guarantees that the vacuum state on the original manifold is mapped to the vacuum state of the new manifold. In other words, the two-point correlation function remains in a form expected for a CFT only when conformal rescaling is done.}. In sum, the mode expansion of $\CO$ in coordinates $(t, \O)$ takes the following form.
\bl
\CO(\t, \O) &= \sum \CO_{nlm}^\dagger \frac{ e^{(\D+2n+l)\t} ~Y_{lm}^\ast (\O) }{M_{nlm}} + \text{h.c.}
\el
Wick rotating to Lorentzian manifold by the substitution $\t = i(1-i\e)t$\cite{Papadodimas:2012aq}, the mode expansion of $\CO$ on Lorentzian $\IR \times S^{d-1}$ that will be used in later sections is obtained.
\bl
\CO(t, \O) &= \sum_{n,l,m} \CO_{nlm} g_{nlm}(t, \O) + \CO_{nlm}^\dagger g_{nlm}^\ast (t, \O) \label{eq:boundmodeexp1}
\\ g_{nlm}(t, \O) &= \frac{1}{M_{nlm}} e^{-i(\D+2n+l)t} ~Y_{lm} (\O)
\el
Two-point functions in Lorentzian signature can be obtained by the same procedure\footnote{The correlation function for the Euclidean signature implicitly contains radial ordering operator, so time ordering operator appears when Wick rotated to Lorentzian signature. The precise procedure is not very different from the procedure outlined in \cite{Papadodimas:2012aq}.}.
\bl
\la T \CO(x) \CO(x') \ra & \propto \frac{1}{\left(\cos(t-t') - \O\cdot \O' + i\e \right)^{\D}} \label{eq:boundary2pt}
\el

\subsection{One-to-one correspondence between free bulk theory and GFT}
It is tempting to conjecture the equivalence of free scalar theory in AdS and GFT defined on its boundary based on the spectrum of energy (conformal dimension for GFT) eigenstates. To claim equivalence, however, requires more nontrivial checks. One of the checks that can be used is to see how correlators of one theory can be related to the correlators of the other. The time-ordered two-point functions $iG(x,x')=\la T \phi(x) \phi(x') \ra$ of AdS free scalar turns out to be \cite{Nastase:adscft}
\bl
iG(x,x') &= \frac{C}{(\cosh^2 s)^{\frac{\D}{2}}}~ {}_2F_1 \left( \frac{\D}{2},\frac{\D+1}{2};\D-\frac{d}{2}+1;\frac{1}{\cosh^2 s}-i\e \right)
\el
where $C$ is an appropriate normalisation constant and $s$ is the geodesic distance, which satisfies the following relation.
\bl
\cosh s &= \frac{\cos(t-t') - \sin\r \sin{\r'} \O \cdot {\O'} }{\cos\r \cos{\r'}} \label{eq:AdSinvdist}
\el
Taking both points to the boundary by the extrapolate map
\bl
\CO(t,\O) &= \lim_{\r \to \pi/2} \frac{\phi(t,\r,\O)}{\cos^\D \r}
\el
the boundary two-point functions become
\bl
\la T \CO(x) \CO(x') \ra & \propto \frac{1}{\left[ \left(\cos(t-t') - \O\cdot \O' \right)^2 + i\e \right]^{\frac{\D}{2}}}
\el
which is equal to \eq{eq:boundary2pt}. Comparing mode expansions \eq{eq:bulkmodeexp1} and \eq{eq:boundmodeexp1}, the natural identification seems to be the following.
\bl
\CO_{nlm} &\leftrightarrow a_{nlm} \label{eq:dictionary1}
\\ \CO_{nlm}^\dagger &\leftrightarrow a_{nlm}^\dagger \label{eq:dictionary2}
\\ g_{nlm}(t, \O) &\leftrightarrow \lim_{\r \to \pi/2} \left( \frac{f_{nlm}(t, \r, \O)}{\cos^\D \r} \right) \label{eq:dictionary3}
\el
The above dictionary reproduces the algebra of mode operators and all correlators of the boundary theory. The normalisation factor $M_{nlm}$ is given by
\bl
M_{nlm}=\frac{N_{nlm}}{P^{(l+d/2-1, \D-d/2)}_n(-1)}
\el

\section{Reconstruction of the bulk} \label{sec:bulkrecon}
Examination of the bulk reconstruction of HKLL\cite{Hamilton:2006az} reveals that bulk reconstruction is a problem of distilling mode operators at the boundary. When boundary mode functions form an orthogonal set with respect to an appropriate inner product defined on the boundary spacetime this problem is readily solved as in the original paper\cite{Hamilton:2006az}. Some subtleties regarding this process is discussed.

\subsection{Review of HKLL reconstruction}
The papers by HKLL attempt to reconstruct the operators of the bulk gravitational theory from the operators of the boundary for the vacuum state dual to pure AdS in the semiclassical limit(corresponding to infinite $N$, infinite 't Hooft coupling $\l$ limit), which is the limit where quasi-local bulk spacetime emerges\cite{Hamilton:2005ju,Hamilton:2006az}. The main idea behind HKLL bulk reconstruction\cite{Hamilton:2006az} is quite simple; find an integration kernel that reads out mode operators of the CFT\footnote{To be more precise, GFT.} operator when integrated against the operator $\CO$ over some boundary integration domain, use the dictionary \eq{eq:dictionary1} and \eq{eq:dictionary2} to substitute the mode operators of the bulk mode expansion \eq{eq:bulkmodeexp1} into integrals over the boundary, and change the order of summation and integration to write a smearing function. Some of the constructions elaborated in \cite{Hamilton:2006az} will be briefly reviewed to demonstrate the procedure and its subtleties.

The global reconstruction of the bulk in \cite{Hamilton:2006az} attempted to reconstruct the bulk operator at the origin. After the smearing function that reconstructs the origin has been constructed, the smearing function for an arbitrary bulk point can be found by using AdS isometries to bring the point to the origin. Restricting the reconstruction to the origin has one notable advantage; since only s-wave components contribute to the bulk operator at the origin, the summation simplifies drastically. The construction starts by dividing the bulk and boundary operators into positive and negative frequency parts.
\bl
\phi(t, \r, \O) &= \phi_{+}(t, \r, \O) + \phi_{-}(t, \r, \O)
\\ \phi_{+}(t, \r, \O) &= \sum_{n,l,m} a_{nlm} f_{nlm}(t, \r, \O) = \left[ \phi_{-}(t, \r, \O) \right]^\dagger
\\ \CO(t, \O) &= \CO_{+}(t, \O) + \CO_{-}(t, \O)
\\ \CO_{+}(t, \O) &= \sum_{n,l,m} \CO_{nlm} g_{nlm}(t, \O) = \left[ \CO_{-}(t, \O) \right]^\dagger
\el
The mode functions $f_{nlm}$ for nonzero $l$ vanish at the origin, so the bulk mode sum simplifies.
\bl
\phi(0) &= \sum_{n} a_{n00} f_{n00}(0) + \text{h.c.}
\\ &= \sum_{n} a_{n00} \frac{Y_{00} ~P_n^{(d/2-1,\D-d/2)}(1)}{N_{n00}} + \text{h.c.} \label{eq:globaloriginmodeexp2}
\el
On the other hand, the mode expansion for the positive frequency part of the boundary operator can be rewritten as follows.
\bl
\CO_{+}(t, \O) &= \sum_{n} \CO_{n00} \frac{Y_{00} ~e^{-i(\D + 2n)t}}{M_{n00}} + \sum_{n, l\neq0, m} \CO_{nlm} g_{nlm}(t, \O)
\el
Using the dictionary \eq{eq:dictionary1}, the mode operators $a_{n00}$ can be read out from the positive frequency part of the boundary operator by the following integral.
\bl
a_{n00} &= \CO_{n00} = \frac{M_{n00}}{\pi}\int_{-\pi/2}^{\pi/2} dt ~e^{i(\D+2n)t} \int \sqrt{g_{\O}} d\O ~Y_{00}^\ast (\O) \CO_{+}(t, \O) \label{eq:boundmoderead1}
\\ &= \int_{-\pi/2}^{\pi/2} dt \int \sqrt{g_{\O}} d\O ~P_{n00}^{+} (t, \O) \CO_{+}(t, \O)
\\ P_{n00}^{+} (t, \O) &= \frac{M_{n00}}{\pi}e^{i(\D+2n)t}~Y_{00}^\ast (\O)
\el
Note that the projection operator $P_{n00}^{+}$ can be always changed to ${P'}_{n00}^{+} = P_{n00}^{+} + \delta P$, where $\delta P$ integrated against $\CO(t,\O)$ always yields zero\footnote{The meaning of this residual freedom is explained in footnote \ref{fn:bdL2}}. This residual freedom in choosing $P_{n00}^{+}$ simplified the summation in \cite{Hamilton:2006az}. Substitution of the above formula into \eq{eq:globaloriginmodeexp2} and changing the order of summation and integration gives the smearing function $K$.
\bl
\phi(0) &= \sum_{n} \frac{Y_{00} ~P_n^{(d/2-1,\D-d/2)}(1)}{N_{n00}} \int_{-\pi/2}^{\pi/2} dt \int \sqrt{g_{\O}} d\O ~P_{n00}^{+} (t, \O) \CO_{+}(t, \O) + \text{h.c.}
\\ &= \left[ \int_{-\pi/2}^{\pi/2} dt \int \sqrt{g_{\O}} d\O ~K_{+}(0 | t, \O) \CO_{+}(t, \O) \right] + \text{h.c.}
\el
In \cite{Hamilton:2006az}, the freedom to change the projection operator ${P}_{n00}^{+} \to P_{n00}^{+} + \delta P$ is used to make $K_{+}$ real, thereby making the smearing function for the negative part $K_{-}$ equal to $K_{+}$. The integration kernel for the positive frequency part and the negative frequency part becomes the same, so the final outcome simplifies to an integration kernel for $\CO(t, \O)$. As the domain of integration at the boundary is the set of points space-like separated from the origin, the global reconstruction uses the boundary data on the points space-like separated from the bulk point of interest. The results of \cite{Hamilton:2006az} are given for future reference.
\bl
K &= \left\{
\begin{aligned}
&\frac{\G(\D-\frac{d}{2}+1) \G(1-d/2)}{\pi \text{vol}(S^{d-1}) \G(\D-d+1)} \lim_{\r \to \pi/2}(2 \s \cos\r)^{\D-d} && \text{even AdS}_{d+1}
\\ &\frac{2 (-1)^{(d/2 - 1)}\G(\D - d/2 + 1)}{\pi \text{vol}(S^{d-1}) \G(\D - d + 1) \G(d/2)} \lim_{\r \to \pi/2} (2 \s \cos\r)^{\D-d}\log (\s \cos\r) && \text{odd AdS}_{d+1}
\end{aligned} \right. \label{eq:HKLLglobal}
\el
The invariant distance $\s$ is given by the formula \eq{eq:AdSinvdist}, where $\s=\cosh s$. As already explained, the smearing function $K$ has support on boundary spacetime regions spacelike separated from the bulk point of interest\footnote{There is a small caveat to this statement. The \emph{domain} of the smearing function $K$ is different from its \emph{support} for a specific bulk point.}. When the bulk point of interest is the origin, the limit $\lim \s \cos\r$ reduces to $\cos t$.

The construction is not very different for AdS-Rindler reconstruction of the bulk in \cite{Hamilton:2006az}. The only caveat is that the summation diverges when order of integration and summation is changed. The divergent sum is made convergent by analytic continuation of coordinates in \cite{Hamilton:2006az}, but as commented by \cite{Morrison:2014jha} this does not seem to be an appropriate way of working with the divergent sum. Rather, it seems more appropriate to give an interpretation of the divergent sum in the context of distribution theory\cite{Morrison:2014jha}. This point of view will be explained in more detail in the following subsection. Another property that is frequently neglected is that the smearing function for the Rindler wedge of boundary domain of dependence $\CD[\CA]$ vanishes completely on the boundary domain of dependence of the complementary boundary subregion $\CD[\CA^c]$, which is obvious from the fact that the domain of smearing function for AdS-Rindler reconstruction is $\CD[\CA]$.

\subsection{Meaning of bulk reconstruction}
Bulk reconstruction aims to obtain bulk data from available boundary data. What is the exact meaning of this statement? Bulk data is encoded in the bulk field $\phi$, but this field itself is not used as an observable; observables are constructed from the bulk field $\phi$ through smearing by integration against a suitable test function $\eta$\cite{Streater:pct}.
\bl
\phi [\eta] &= \int d^{d+1}x \sqrt{\abs{g}} \eta(x) \phi(x)
\el
The same applies to observables of the boundary theory\cite{Morrison:2014jha}.
\bl
\CO [\z] &= \int d^{d}Y \sqrt{\abs{\g}} \z(Y) \CO(Y)
\el
In both theories, finite sums of finite products of the above smeared fields define the algebra of local observables. When smearing function $K$ is introduced to interpolate the formulas, the following relation is obtained.
\bl
\phi [\eta] &= \int d^{d+1}x \sqrt{\abs{g}} \eta(x) \phi(x)
\\ &= \int d^{d+1}x \sqrt{\abs{g}} \eta(x) \int d^{d}Y \sqrt{\abs{\g}} K(x|Y) \CO(Y)
\\ &= \int d^{d}Y \sqrt{\abs{\g}} \left[ \int d^{d+1}x \sqrt{\abs{g}} \eta(x) K(x|Y) \right] \CO(Y)
\\ &= \CO [\eta^\p]
\el
The test function for the boundary $\eta^\p$ is defined as follows.
\bl
\eta^\p (Y) &= \int d^{d+1}x \sqrt{\abs{g}} \eta(x) K(x|Y)
\el
This is how \cite{Morrison:2014jha} argued that the problematic divergent behaviour of mode summation in AdS-Rindler reconstruction is actually not a problem. This relation suggests that the smearing function $K$ should rather be considered as a rule to assign bulk test functions to their boundary counterparts, not as a rule that relates bulk field values to their boundary counterparts.

\subsection{Distilling mode operators and constructing the smearing function} \label{sec:boundprojdef}
The gist of bulk reconstruction lies in obtaining the mode operators at the boundary. The role of the smearing function $K$ is to automatise the procedure of obtaining a mode operator through a simple integral from the boundary, attaching the corresponding mode function to the mode operator, and summing the result over all mode operators. The problem is that in some cases, it is hard to distill the wanted mode operators by a simple integral. When mode functions of the boundary field are orthogonal, the wanted mode operator can be regained by exploiting this orthogonality. Unfortunately, such a miracle will not happen generally; this is why \cite{Hamilton:2006az} needed to separate positive frequency modes and negative frequency modes in the beginning. This subsection is devoted to working around this nonorthogonality problem.

Suppose that a region $\Xi$ of boundary spacetime is given as the domain for the wanted smearing function $K$. In the case of HKLL reconstruction\cite{Hamilton:2006az}, the domain of $K$ is taken to be the whole boundary spacetime for the global reconstruction and boundary of the AdS-Rindler wedge for the AdS-Rindler reconstruction. The mode functions ${g}_{nlm}$ and ${g}_{nlm}^\ast$ constitute a \emph{complete} set of basis\footnote{A complete set of basis for functions satisfying the boundary equation of motion defined by boundary Hamiltonian, which is the dilatation operator. The space of such functions only covers a subspace $V_1$ of the full boundary function space $V$ of $L^2$ norm. The residual freedom of shifting the projection operator $P^{+} \to P^{+} + \delta P$ is equivalent to adding a projection operator for a vector in $V_1^\perp$, the orthogonal complement of $V_1$. This freedom is interpreted as a kind of gauge freedom in \cite{Mintun:2015qda} and \cite{Freivogel:2016zsb}.\label{fn:bdL2}}, while it is hard to find an inner product on the boundary that makes them orthonormal. Nevertheless, it is formally possible to discern contributions from different mode functions. Define the following inner product on the boundary spacetime region $\Xi$.
\bl
(f,g)_\Xi &\equiv \int_\Xi f^\ast g ~ dV
\el
$dV = \sqrt{g_{\O}} dt d\O$ is the standard spacetime volume measure of the boundary. The mode functions form a countably infinite basis, which is schematically refered to as $g_m$. Construct the Gram matrix $\bold{g} = g_{mn} $ by the given inner product.
\bl
g_{mn} &\equiv (g_m, g_n)_\Xi
\el
The Gram matrix is Hermitian, i.e. $\bold{g}^\dagger = \bold{g}$. Suppose that the inverse of $\bold{g}$, $\bold{g}^{-1} = g^{mn}$, exists. The inverse is defined by the relation $g^{mn} g_{nl} = g_{ln} g^{nm} = \delta_l^m$. The inverse can be used to define projection operator $P^m$.
\bl
P^m (f) &\equiv \int_\Xi g^{mn} g_n^\ast f ~dV
\\ P^m(g_n) &= g^{ml} \int_\Xi g_l^\ast g_n ~dV = g^{ml} g_{ln} = \delta^m_n
\el
The mode operators $\CO_{nlm}$ can be extracted from the given boundary data on $\S$ by this projection operator $P^{nlm}$. For convenience, schematic mode index $m$ is used below.
\bl
\CO_{nlm} &= N_{nlm} P^{nlm}(\tilde{\CO}(t,\O)) = N_{nlm} g^{nlm,m'} \int_\Xi g_{m'}^\ast \tilde{\CO}(t,\O) ~dV
\\ \CO_{nlm}^\dagger &= N_{nlm} P^{nlm\ast}(\tilde{\CO}(t,\O)) = N_{nlm} g^{nlm\ast,m'} \int_\Xi g_{m'}^\ast \tilde{\CO}(t,\O) ~dV
\el
The asterisk on $P^{nlm\ast}$ is intended as a reminder that this operator projects onto $g_{nlm}^\ast$. Although this is an unfortunate notation that may cause confusion, the integration kernel for extracting the mode operator will be simplified to ${P}_{nlm}^{\pm}$, ${P}_{nlm}^{+}$ for $\CO_{nlm}$ and ${P}_{nlm}^{-}$ for $\CO_{nlm}^\dagger$. Given projection operators, obtaining the smearing function is a trivial task.
\bl
\tilde{\phi}(t, \r, \O) &= \sum a_{nlm} f_{nlm}(t, \r, \O) + a_{nlm}^\dagger f_{nlm}^\ast(t, \r, \O)
\\ &= \sum f_{nlm}(t, \r, \O) \int_\Xi {P}_{nlm}^{+} (t', \O') \tilde{\CO}(t',\O') ~dV' \nn
\\ & \hskip 10pt + \sum f_{nlm}^\ast (t, \r, \O) \int_\Xi {P}_{nlm}^{-} (t', \O') \tilde{\CO}(t',\O') ~dV' \nn
\\ &= \sum f_{nlm}(t, \r, \O) N_{nlm} g^{nlm,m'} \int_\Xi g_{m'}^\ast \tilde{\CO}(t',\O') ~dV' \nn
\\ & \hskip 10pt + \sum f_{nlm}^\ast (t, \r, \O) N_{nlm} g^{nlm\ast,m'} \int_\Xi g_{m'}^\ast \tilde{\CO}(t',\O') ~dV' \nn
\\ &= \int_\Xi K(t, \r, \O | t', \O') \tilde{\CO}(t',\O') dV'
\el
The explicit expression for the smearing function $K(t, \r, \O | t', \O')$ is as follows. Note that this smearing function has support on the whole domain, where domain is taken to be the region $\Xi$ of the boundary spacetime.
\bl
K(t, \r, \O | t', \O') &= \sum_{n,l,m} f_{nlm}(t, \r, \O) {P}_{nlm}^{+} (t', \O') + f_{nlm}^\ast (t, \r, \O) {P}_{nlm}^{-} (t', \O')
\\ &= \sum_{n,l,m,m'} N_{nlm} \left( f_{nlm}(t, \r, \O) g^{nlm,m'} + f_{nlm}^\ast (t, \r, \O) g^{nlm\ast,m'} \right) g_{m'}^\ast(t',\O')
\el

\subsection{Covariance under mode function choices}
A natural question is to ask whether the smearing function constructed above depends on the choice of mode functions, which form a basis of the solution space of the wave equation. Some algebra shows that it does not depend on any specific choice: An example is given in \cite{Hamilton:2006az}, where it is shown that smearing functions constructed from global patch mode functions and Poincar\'e patch mode functions are equivalent up to some irrelevant factor. Having a schematic form of the smearing function is helpful for showing this independence. Schematically, the smearing function constructed can be written as follows.
\bl
K(x|Y) &= \sum f_{k}(x) g^{kl} g_{l}^\ast (Y)
\\ g_{ij} &= \int_{\Xi} dY g_{i}^\ast (Y) g_{j}(Y), ~~ g^{ki}g_{ij} = \delta^{k}_{j}
\el
The labels refer to schematic mode indices, i.e. the indices run over mode functions and their complex conjugates collectively. Suppose that another set of mode functions, ${f'}_{k}(x)$, is given. The smearing function constructed from this set is written in the following way.
\bl
{K'}(x|Y) &= \sum {f'}_{k}(x) {g'}^{kl} {g'}_{l}^\ast (Y)
\\ {g'}_{ij} &= \int_{\Xi} dY {g'}_{i}^\ast (Y) {g'}_{j}(Y), ~~ {g'}^{ki}{g'}_{ij} = \delta^{k}_{j}
\el
The mode functions form a basis, so there exists a matrix that relates the different bases.
\bl
f_{k}(x) &= \sum_{l} \a (k|l) {f'}_{l}
\\ {f'}_{k}(x) &= \sum_{l} \b (k|l) {f}_{l}
\el
The matrices $\a$ and $\b$ are inverses of each other; $\sum \a(i|j) \b(j|k) = \sum \b(i|j) \a(j|k) = \delta_{ik}$. The sets of boundary mode functions $g_{k}$ and ${g'}_{k}$ follow the same relations, since boundary mode functions are obtained as some limit of the bulk mode functions. A bit of algebra can show that the relation $K(x|Y) = {K'}(x|Y)$ holds.
\bl
{K'}(x|Y) &= \sum {f'}_{k}(x) {g'}^{kl} {g'}_{l}^\ast (Y)
\\ &= \sum {f'}_{k}(x) {[\b^\ast(i|l) g_{ij} \b(j|k) ]}^{-1} \b^\ast (l|m) {g}_{m}^\ast (Y) \nn
\\ &= \sum {f'}_{k}(x) \a (j|k) g^{ji} \a^\ast (i|l) \b^\ast (l|m) {g}_{m}^\ast (Y) \nn
\\ &= \sum \a (j|k) {f'}_{k}(x)  g^{jl} {g}_{l}^\ast (Y) \nn
\\ &= \sum {f}_{k}(x)  g^{kl} {g}_{l}^\ast (Y) = K(x|Y)
\el
This relation implies that the construction of the smearing function is independent of the choice of mode functions.

\subsection{Feasibility of distillation} \label{sec:feasibility}
Since the Gram matrix $\bold{g}$ is an infinite dimensional matrix, it is not clear whether it is possible to calculate the inverse matrix elements or not. However, it is possible to write an formal expression that corresponds to $\bold{g}^{-1}$. Decompose $\bold{g}$ into its diagonal and off-diagonal parts $\bold{D}$ and $\bold{h}$. $\bold{D}^{-1}$ is easily calculated since none of the diagonal components of $\bold{g}$ are zero. Then $\bold{g}^{-1}$ has the following formal expression.
\bl
\bold{g} &= \bold{D} - \bold{h}
\\ \bold{g}^{-1} &= \bold{D}^{-1}( \iden + \bar{\bold{h}} + \bar{\bold{h}}^2 + \bar{\bold{h}}^3 + \cdots)
\\ \bar{\bold{h}} &= \bold{h}\bold{D}^{-1}
\el
If the spectrum $\l_m$ of $\bar{\bold{h}}$ satisfies the criteria $\forall m\{ \abs{\l_m} < 1 \}$ then this formal expression is exact. Unfortunately, there is no good criteria for determining the upper bound for the spectrum of $\bar{\bold{h}}$, so invertibility of $\bold{g}$ is not guaranteed.

On the other hand, noninvertibility of $\bold{g}$ seems natural for reasons given in section \ref{sec:ewreconsubsec}. A possible resolution is to consider the theory on both sides of the correspondence to be an effective description. Introducing a cut-off to $n$ and $l$ restricts the dimension of the vector space generated by mode functions to be finite and $\bold{g}$ becomes a finite dimensional matrix, which is always invertible. Considering that GFT is only an effective description of a CFT\cite{ElShowk:2011ag}, this seems to be a natural resolution to the noninvertibility problem.

\section{Restriction to subregions of the bulk} \label{sec:bulksubregion}
The full Hilbert space $\CH$ of the bulk can be decomposed into a product of subregion Hilbert spaces $\CH_\a$. These Hilbert spaces can be constructed as Fock spaces of respective mode operators, and it is possible to explicitly construct the relations between mode operators of subregions and mode operators of the full space which are a generalised version of Bogoliubov transformations. Since it is quite cumbersome to work with subregion mode operators, \emph{operator truncation} and \emph{mode function resummation} will be introduced to work with full Hilbert space mode operators instead.

\subsection{Separation of Hilbert space into subregion Hilbert spaces}
Conventionally, the Hilbert space of a QFT is constructed as the Fock space obtained from creation and annihilation operators. Creation and annihilation operators are obtained from a mode function decomposition of the corresponding operator, and the inner product \eq{eq:fbraketdef} defined between functions can be used to explicitly decompose the operator.
\bl
\phi (t, x) &= \sum_k a_k f_k(t, x) + a_k^\dagger f_k^\ast (t, x)
\\ a_k &= \fbraket{ f_k }{ \phi }_\S
\\ a_k^\dagger &= -\fbraket{ f_k^\ast}{ \phi }_\S
\el
$k$ denotes the set of indices, discrete or continuous, that characterises each distict mode of a complete set. The mode functions are assumed to satisfy the conventional normalisation $\fbraket{f_k}{f_{k'}}_\S = -\fbraket{f_k^\ast}{f_{k'}^\ast}_\S = \delta_{kk'}$.

A similar construction can be used to build the Hilbert spaces for subregions. A state in the Hilbert space $\CH$ of a QFT is determined by the field configuration on a time slice $\S$. If the time slice $\S$ is decomposed into a discrete set of subregions $\S_\a$, the field configuration on the subregion $\S_\a$ determines a state in the subregion Hilbert space $\CH_\a$. The tensor product of subregion Hilbert spaces becomes the total Hilbert space, i.e. $\CH=\prod \CH_\a$. The complete set of basis for a subregion Hilbert space $\CH_\a$ can be constructed in an analogous way to the full Hilbert space $\CH$; find the normalised subregion mode functions $f_{\a | k}$ and $f_{\a | k}^\ast$, use the inner product modified for subregions $\fbraket{f}{g}_{\S_\a}$ to find the creation and annihilation operators for subregion $\S_\a$, and use the mode operators to construct the Fock states for subregion $\S_\a$.

Having a new coordinate patch simplifies finding the normalised subregion mode functions, at least conceptually. Given a subregion $\S_\a$ of the time slice $\S$, it is possible to find a coordinate patch $U_\a$ that only covers the bulk domain of dependence $\CD_B[\S_\a]$\footnote{$\CD_B[\S_\a]$ is a set of bulk points which any fully extended causal curve passing through the point must intersect $\S_\a$.}. The subregion mode functions are solutions to the wave equation that only has support on the coordinate patch $U_\a$ and regions of the full spacetime causally connected to $\S_\a$; in other words, subregion mode functions for the subregion $\S_\a$ are solutions to the wave equation that does not have any support on coordinate patch(es) $U_\b$, $\b \neq \a$. Solving the wave equation on $U_\a$ and finding a complete set of solutions with Dirichlet boundary conditions\footnote{Treating the fields as operator valued \emph{distributions} play a subtle role here as Dirichlet boundary conditions to subregion mode functions implies that $\phi (x)$ at $x \in \p\S_\a$ for some $\a$ is always zero, where $\p\S_\a$ denotes the boundary of $\S_\a$. In the context of distribution theory, however, this does not matter since the set $\p\S_\a$ constitutes a measure zero set and operators are obtained by integrating the field $\phi$ with respect to the corresponding test functions.} give the subregion mode functions $f_{\a | k}$ and $f_{\a | k}^\ast$.

The inner product between subregion mode functions can be defined in an analogous way.
\bl
\fbraket{f}{g}_{\S_\a} &\equiv i \int_{\S_\a} \hskip -5pt \sqrt{h} dy ~ n^\m \left( f^\ast \del_\m g- g\del_\m f^\ast \right)
\el
$\S_\a$ denotes the subregion of the time slice, the spacelike surface on which the inner product is evaluated. Other symbols are defined in the same way as in \eq{eq:fbraketdef}. The field $\phi$ has an expansion in subregion mode functions and subregion mode operators, and subregion mode operators admit an expression in terms of subregion inner products.
\bl
\phi (t, x) &= \sum_{\a, k} a_{\a | k} f_{\a | k}(t_\a, x_\a) + a_{\a | k}^\dagger f_{\a | k}^\ast(t_\a, x_\a)
\\ a_{\a | k} &= \fbraket{  f_{\a | k} }{ \phi }_{\S_\a}
\\ a_{\a | k}^\dagger &= -\fbraket{  f_{\a | k}^\ast}{ \phi }_{\S_\a}
\el
The coordinates $(t_\a, x_\a)$ refers to the coordinate labels of the coordinate patch $U_\a$; the coordinate patch that covers $\CD_B[\S_\a]$.

The Fock states or the single particle states that are used as the basis for constructing the subregion Hilbert space can be constructed in the same way as how Fock states for the full Hilbert space in conventional QFT is constructed, so it will not be elaborated here. Having constructed the Fock states for each subregion Hilbert space $\CH_\a$, the bases for the full Hilbert space $\CH$ can be constructed as tensor products of subregion Fock states.

\subsection{Relationships between mode operators}
It turns out that the mode operators $a_k$ and $a_k^\dagger$ of the full Hilbert space can be expressed as a linear combination of subregion mode operators $a_{\a | k}$ and $a_{\a | k}^\dagger$, and vice versa. The relation can be obtained from two distinct but equivalent mode decomposition of the field $\phi (x,t) $.
\bl
\phi (t, x) &= \sum_k a_k f_k(t, x) + a_k^\dagger f_k^\ast (t, x)
\\ &= \sum_{\a, k} a_{\a | k} f_{\a | k}(t_\a, x_\a) + a_{\a | k}^\dagger f_{\a | k}^\ast(t_\a, x_\a)
\el
It is more instructive to write the mode expansion in the following form, which is motivated by the fact that mode functions can be considered as vectors of a vector space.
\bl
\phi (t, x) &= \sum_k a_k \fket{ f_k } + a_k^\dagger \fket{ f_k^\ast }
\\ &= \sum_{\a, k} a_{\a | k} \fket{ f_{\a | k} } + a_{\a | k}^\dagger \fket{ f_{\a | k}^\ast }
\el
Dual vectors $\fbra{f}$ are defined as functionals that map functions that satisfy the wave equation $\fket{g}$ to numbers $\fbraket{f}{g}$. Generalised Bogoliubov relations relating $a_{k}$ and $a_{k}^\dagger$ to $a_{\a | k}$ and $a_{\a | k}^\dagger$ can be read out from the above expression using the inner product between mode functions.
\bl
a_{k} & = \fbraket{ f_{k} }{\phi}_\S = \sum_{\a, q} \fbraket{ f_{k}}{f_{\a | q}}_{\S_\a} a_{\a | q} + \fbraket{ f_{k} }{ f_{\a | q}^\ast}_{\S_\a} a_{\a | q}^\dagger
\\ a_{k}^\dagger & = - \fbraket{ f_{k}^\ast }{\phi}_\S = \sum_{\a, q} - \fbraket{ f_{k}^\ast }{f_{\a | q}}_{\S_\a} a_{\a | q} - \fbraket{ f_{k}^\ast }{ f_{\a | q}^\ast}_{\S_\a} a_{\a | q}^\dagger
\\ a_{\a | k} & = \fbraket{ f_{\a | k} }{\phi}_{\S_\a} = \sum_q \fbraket{ f_{\a | k}}{f_{q}}_{\S_\a} a_{q} + \fbraket{ f_{\a | k}}{f_{q}^\ast}_{\S_\a} a_{q}^\dagger
\\ a_{\a | k}^\dagger & = - \fbraket{ f_{\a | k}^\ast }{\phi}_{\S_\a} = \sum_q - \fbraket{ f_{\a | k}^\ast}{f_{q}}_{\S_\a} a_{q} - \fbraket{ f_{\a | k}^\ast }{f_{q}^\ast}_{\S_\a} a_{q}^\dagger
\el
Using inner product relations, the above expression can be recast in a more familiar form.
\bl
a_{k} & = \sum_{\a, q} \fbraket{ f_{k}}{f_{\a | q}}_{\S_\a} a_{\a | q} + \fbraket{ f_{k} }{ f_{\a | q}^\ast}_{\S_\a} a_{\a | q}^\dagger
\\ a_{k}^\dagger & = \sum_{\a, q} \left[ \fbraket{ f_{k}}{f_{\a | q}}_{\S_\a} \right]^\ast a_{\a | q}^\dagger + \left[ \fbraket{ f_{k} }{ f_{\a | q}^\ast}_{\S_\a} \right]^\ast a_{\a | q}
\\ a_{\a | k} & = \sum_q \left[ \fbraket{ f_{q}}{f_{\a | k}}_{\S_\a} \right]^\ast a_{q} - \fbraket{ f_{q} }{ f_{\a | k}^\ast}_{\S_\a} a_{q}^\dagger
\\ a_{\a | k}^\dagger & = \sum_q \fbraket{ f_{q}}{f_{\a | k}}_{\S_\a} a_{q}^\dagger - \left[ \fbraket{ f_{q} }{ f_{\a | k}^\ast}_{\S_\a} \right]^\ast a_{q}
\el
In terms of the usual generalised Bogoliubov coefficients $\a$ and $\b$, the coefficient $\fbraket{ f_{q}}{f_{\a | k}}_{\S_\a}$ corresponds to the coefficient $\a (\a, k | q)$ and the coefficient $\fbraket{ f_{q} }{ f_{\a | k}^\ast}_{\S_\a}$ corresponds to the coefficient $\b^\ast (\a, k | q)$. The analogues of the unitarity condition $\abs{\a}^2 - \abs{\b}^2 = 1$ are the following two relations.
\bl
\delta_{kk'} &= \sum_{\a, q} \a (\a, q | k) \a^\ast (\a, q | k') - \b^\ast (\a, q | k) \b (\a, q | k')
\\ \delta_{\a \a'}\delta_{kk'} &= \sum_{q} \a^\ast (\a, k | q) \a (\a', k' | q) - \b^\ast (\a, k | q) \b (\a', k' | q)
\el
The analogues of the unitarity condition $\a\b - \b\a = 0$ are the following two relations.
\bl
0 &= \sum_{\a, q} \a (\a, q | k) \b^\ast (\a, q | k') - \b^\ast (\a, q | k) \a (\a, q | k')
\\ 0 &= \sum_{q} \a (\a, k | q) \b (\a', k' | q) - \b (\a, k | q) \a (\a', k' | q)
\el
After a bit of algebra, the bra-ket notation for mode functions yields equivalent but more intuitive expressions.
\bl
\delta_{kk'} &= 
- \fbra{ f_k^\ast } \left[ \sum_{\a, q} \fket{ f_{\a | q} } \fbra{ f_{\a | q} } - \fket{ f_{\a | q}^\ast } \fbra{ f_{\a | q}^\ast } \right] \fket{ f_{k'}^\ast }
\\ \delta_{\a \a'}\delta_{kk'} &= 
- \fbra{ f_{\a | k}^\ast } \left[ \sum_{q} \fket{ f_{q} } \fbra{ f_{q} } - \fket{ f_{q}^\ast } \fbra{ f_{q}^\ast } \right] \fket{ f_{\a' | k'}^\ast }
\\ 0 &= - \fbra{ f_k } \left[ \sum_{\a, q} \fket{ f_{\a | q} } \fbra{ f_{\a | q} } - \fket{ f_{\a | q}^\ast } \fbra{ f_{\a | q}^\ast } \right] \fket{ f_{k'}^\ast }
\\ 0 &= - \fbra{ f_{\a | k}^\ast } \left[ \sum_{q} \fket{ f_{q} } \fbra{ f_{q} } - \fket{ f_{q}^\ast } \fbra{ f_{q}^\ast } \right] \fket{ f_{\a' | k'} }
\el
The meaning is clear; the sum $\sum \fket{ f_{k} } \fbra{ f_{k} } - \fket{ f_{k}^\ast } \fbra{ f_{k}^\ast }$ over an index of a complete set $k$ is nothing but the identity.

\subsection{Operator truncation and mode function resummation}
Suppose that subregion $\S_\b$ is not available. The state is now better described by the reduced density matrix $ \r = \text{Tr}_{\CH_\b} \ket{\psi}\bra{\psi} $ and only the operators that do not act on $\CH_\b$ are considered. In applications to QFT, however, obtaining the coefficients for reduced density matrices is a daunting task if not impossible. Therefore it is more desirable to work with the state $\ket{\psi}$ rather than the reduced density matrix. This is the motivation for developing \emph{operator truncation}.

For simplicity, assume that the total Hilbert space $\CH$ is a tensor product of two subsystem Hilbert spaces $\CH_1$ and $\CH_2$. Also, assume that all operators of interest admit the following decomposition,
\bl
O = O_1\otimes \iden_2 + \iden_1 \otimes O_2 \label{eq:localdecomp}
\el
where $\iden_1$ and $\iden_2$ are identity operators of $\CH_1$ and $\CH_2$, $O_1: \CH_1 \to \CH_1$ is an operator of $\CH_1$, and $O_2: \CH_2 \to \CH_2$ is an operator of $\CH_2$, respectively. The class of operators having a decomposition of the form \eq{eq:localdecomp} consists of linear combination of local operators.\footnote{In quantum information language, the class of operators having the form $O_1\otimes \iden_2$ or $\iden_1 \otimes O_2$ are called local operations \cite{Diosi:info}.}

Suppose subsystem 2 is inaccessible. The information that can be withdrawn is encoded in the reduced density matrix $\r_1=\text{Tr}_{\CH_2}\ket{\psi}\bra{\psi}$. In this case, the natural modification to an operator $O : \CH \to \CH$ would be to trace it over $\CH_2$ and divide by $N_2=\text{dim }\CH_2$.
\bl
\ket{\psi} &\quad \to \quad \r_1=\text{Tr}_{\CH_2}\ket{\psi}\bra{\psi}
\\ O &\quad \to \quad O_{\text{red}}= \frac{1}{N_2} \text{Tr}_{\CH_2}O
\el
If the operator $O$ is of the form \eq{eq:localdecomp} and $\text{Tr}_{\CH}O=0$\footnote{This condition can be always met by substracting an appropriate multiple of identity from $O$.} holds, something more interesting can be said. Define $\tilde{O}$, the \emph{truncation of $O$}, as follows.
\bl
O = O_1\otimes \iden_2 + \iden_1 \otimes O_2  \quad  \to  \quad  \tilde{O} = O_1 \otimes \iden_2 \label{eq:truncation}
\el
$\tilde{O} = O_{\text{red}}\otimes \iden_2$ up to a multiple of identity because $\text{Tr}_{\CH}O=0$. If any of $\text{Tr}_{\CH_1}O_1=0$ or $\text{Tr}_{\CH_2}O_2=0$ holds together with $\text{Tr}_{\CH}O=0$, then the equality is exact. Ignoring the multiple of identity, the following relation holds as well.
\bl
\text{Tr}_{\CH_1} \left[ O_{\text{red}} \r_1 \right] = \text{Tr}_{\CH_1}\left[ O_1\r_1 \right] = \bra{\psi} \tilde{O} \ket{\psi} \label{eq:truncapp}
\el
The formula \eq{eq:truncapp} suggests that through appropriate truncation of operators, it is possible to work with the ground state instead of the reduced density matrix.

Truncation defined above can be readily generalised to free scalar theory in AdS, since the full operator is already in a form similar to \eq{eq:localdecomp}.
\bl
\phi (t, x) &= \sum_{\a, k} a_{\a | k} \fket{ f_{\a | k} } + a_{\a | k}^\dagger \fket{ f_{\a | k}^\ast }
\\ &= \sum_{\a \in A, k} \left[ a_{\a | k} \fket{ f_{\a | k} } + a_{\a | k}^\dagger \fket{ f_{\a | k}^\ast } \right] + \sum_{\a \notin A, k} \left[ a_{\a | k} \fket{ f_{\a | k} } + a_{\a | k}^\dagger \fket{ f_{\a | k}^\ast } \right]
\el
Considering $A$ as the set of accessible subregions, the truncated operator $\tilde{\phi}$ has the following mode expansion.
\bl
\tilde{\phi}(t, x) &= \sum_{\a \in A, k} a_{\a | k} ~ \fket{ f_{\a | k} } + a_{\a | k}^\dagger ~ \fket{ f_{\a | k}^\ast }
\el
A little algebra using generalised Bogoliubov transforms reveals that $\tilde{\phi}$ can be expanded by full Hilbert space mode operators $a_{k}$ and $a_{k}^\dagger$.
\bl
\tilde{\phi}(t, x) &= \sum_{\a \in A, k} a_{k} \left[ \iden_\a \fket{ f_{k} } \right] + a_{k}^\dagger \left[\iden_\a \fket{ f_{k}^\ast } \right]
\\ & = \sum_{k} a_{k} ~ \fket{ \tilde{f}_{k} } + a_{k}^\dagger ~ \fket{ \tilde{f}_{k}^\ast } \label{eq:truncmodeexp}
\\ \iden_\a & \equiv \sum_{k} \fket{f_{\a | k}}\fbra{f_{\a | k}} - \fket{f_{\a | k}^\ast}\fbra{f_{\a | k}^\ast}
\\ \fket{ \tilde{f}_{k} } & \equiv \sum_{\a \in A} \iden_\a \fket{ f_{k} } =\sum_{\a \in A, k'} \fket{f_{k'}} \fbraket{f_{k'}}{ f_{k} }_{\S_\a} - \fket{f_{k'}^\ast} \fbraket{f_{k'}^\ast}{ f_{k} }_{\S_\a}
\el
It is worthy of note that $\iden_\a$ acts as an projection operator, since the mode functions $f_{\a | k}$ and $f_{\a | k}^\ast$ are assumed to constitute a complete basis for the solution space of the wave equation in domain of dependence of subregion $\a$. When this operator is sandwiched in between the inner product, it restricts the region of integration to $\S_\a$.
\bl
\fbraket{f}{g}_\S & = i \int_{\S} \hskip -3pt \sqrt{h} dy ~ n^\m \left( f^\ast \del_\m g- g\del_\m f^\ast \right)
\\ \fbra{f} \iden_\a \fket{g}_\S & = i \int_{\S_\a} \hskip -5pt \sqrt{h} dy ~ n^\m \left( f^\ast \del_\m g- g\del_\m f^\ast \right)
\el
This means resummed mode functions $\tilde{f}_{k}$ and $\tilde{f}_{k}^\ast$ can be evaluated without directly working out the solutions to the equation of motion for subregion coordinate patches. Note that in $\CD_B[\cup_{\a \in A} \S_\a]$ the resummed mode function $\tilde{f}_k$ and the original mode function $f_k$ have the same value. This property will play a critical role in the following section.

\section{Bulk reconstruction on the entanglement wedge} \label{sec:entwedgerecon}
The algorithm for constructing the bulk from boundary subregions will be described. After briefly reviewing the recent proof\cite{Dong:2016eik} that the dual bulk region is the entanglement wedge, operator truncation for the boundary will be discussed. The bulk reconstruction algorithm will follow from combining the results of all previous sections. Some comments on covariance of the construction are made as a final remark.

\subsection{Review on reconstructibility of the entanglement wedge}
To understand the proof of entanglement wedge reconstruction given by DHW\cite{Dong:2016eik}\footnote{\cite{Bao:2016skw} gives another approach to the subject, albeit restricted to pure AdS space.}, understanding of the viewpoint put forward by ADH is essential; viewing AdS/CFT correspondence as having quantum error correcting code-like structure\cite{Almheiri:2014lwa}. In ADH viewpoint, the bulk Hilbert space $\CH_{bulk}$ is viewed as a code subspace $\CH_{code}$ of the full boundary Hilbert space $\CH_{CFT}$. A bulk operator $\CO_b : \CH_{bulk} \to \CH_{bulk}$ can be realised by a boundary operator $\CO : \CH_{CFT} \to \CH_{CFT}$ such that the following relation holds for any state $\ket{\psi} \in \CH_{bulk} = \CH_{code}$.
\bl
\CO_b \ket{\psi} &= \CO \ket{\psi} \label{eq:reconrel1}
\\ \CO_b^\dagger \ket{\psi} &= \CO^\dagger \ket{\psi} \label{eq:reconrel2}
\el

The error correcting code-like structure appears when operators that acts only on a subspace of the full Hilbert space are considered. Suppose that the bulk and boundary Hilbert space factorises into $\CH_{bulk} = \CH_{a} \otimes \CH_{\bar{a}}$ and $\CH_{CFT} = \CH_{A} \otimes \CH_{\bar{A}}$. Consider a bulk operator that acts on the factorised part of the bulk Hilbert space $O_b : \CH_{a} \to \CH_{a}$ that can be realised by a boundary operator $O : \CH_{A} \to \CH_{A}$, meaning that the analogues of equations \eq{eq:reconrel1} and \eq{eq:reconrel2} hold for $O_b$ and $O$. This is equivalent to the statement that for any operator $X_{\bar{A}} : \CH_{\bar{A}} \to \CH_{\bar{A}}$ and any states $\ket{\psi}, \ket{\phi} \in \CH_{bulk}$ the following relation holds\cite{Almheiri:2014lwa}.
\bl
\bra{\psi} \left[ O_b, X_{\bar{A}} \right] \ket{\phi} = 0 \label{eq:reconrel3}
\el

There is some arbitrariness in the choice of the boundary factorisation $\CH_{CFT} = \CH_{A} \otimes \CH_{\bar{A}}$, which depends on the choice of the bulk factorisation $\CH_{bulk} = \CH_{a} \otimes \CH_{\bar{a}}$\footnote{This arbitrariness is greatest when the factor Hilbert space of the bulk $\CH_{a}$ is taken to be the Hilbert space generated by states obtained from the vacuum by acting operators localised near the centre of the AdS. This is why \cite{Almheiri:2014lwa} claims that this choice is the most robust code subspace against arbitrary boundary erasures.}. This arbitrariness can be exploited to realise the operator $O_b$ by $O$ which does not act on the inaccessible part $\CH_{\bar{A}}$ of the full Hilbert space, so that any error that occurred on $\CH_{\bar{A}}$ does not affect the action of the operator $O_b$. When a different error occurred so that a different part $\CH_{\bar{A'}}$ of the full Hilbert space $\CH_{CFT} = \CH_{A'} \otimes \CH_{\bar{A'}}$ is inaccessible, a different realisation $O' : \CH_{A'} \to \CH_{A'}$ can be used to realise the same operator $O_b$. This resiliency in arbitrary erasure of the boundary is at the heart of the ADH proposal\cite{Almheiri:2014lwa}.

The DHW argument\cite{Dong:2016eik} relies on the observation of JLMS that relative entropies computed in the bulk and the boundary coincide\cite{Jafferis:2015del}. Decompose the boundary Hilbert space by $\CH_{CFT} = \CH_{A} \otimes \CH_{\bar{A}}$, and decompose the bulk Hilbert space by $\CH_{bulk} = \CH_{a} \otimes \CH_{\bar{a}}$ where $\CH_{a}$ denotes the Hilbert space of bulk excitations in the entanglement wedge $\CE[A]$\footnote{Entanglement wedge of $A$ is defined as bulk domain of dependence of any bulk spacelike surface having $A$ and its HRT surface $\chi_A$ as its boundary. The HRT surface $\chi_A$ is defined as a codimension 2 bulk surface of extremal area (having the least area if multiple of such surfaces exist) homologous to $A$ which shares its boundary, $\p \chi_A = \p A$\cite{Dong:2016eik}.} of boundary region $A$. Likewise, $\CH_{\bar{a}}$ denotes the Hilbert space of bulk excitations in $\CE[\bar{A}]$. The JLMS result\cite{Jafferis:2015del} can be formulated as follows, which is the form required by the DHW argument.
\bl
S(\r_{\bar{A}}|\s_{\bar{A}}) = S(\r_{\bar{a}}|\s_{\bar{a}}) \label{eq:JLMS1}
\el
The left hand side gives the relative entropy of the boundary subregion theory, while the right hand side gives the relative entropy of the bulk theory \emph{inside the entanglement wedge} of the given boundary subregion. Since relative entropy $S(\r|\s) = 0$ if and only if $\r=\s$, the following conclusion can be derived from the JLMS result.
\bl
\r_{\bar{a}}=\s_{\bar{a}} \imp \r_{\bar{A}} = \s_{\bar{A}} \label{eq:JLMS2}
\el
What has been shown by DHW\cite{Dong:2016eik} is that \eq{eq:JLMS2} implies \eq{eq:reconrel3} for density matrices formed by partial trace over pure states. To be specific, the density matrices $\r_{\bar{A}}$, $\s_{\bar{A}}$, $\r_{\bar{a}}$, and $\s_{\bar{a}}$ are assumed to be obtained by the following partial traces.
\bl
\begin{aligned}
&\r_{\bar{A}} \equiv \text{Tr}_A \ket{\phi}\bra{\phi}, && \s_{\bar{A}} \equiv \text{Tr}_A \ket{\psi}\bra{\psi} 
\\ &\r_{\bar{a}} \equiv \text{Tr}_a \ket{\phi}\bra{\phi}, && \s_{\bar{a}} \equiv \text{Tr}_a \ket{\psi}\bra{\psi}
\end{aligned}
\el
This in turn implies that equivalents of \eq{eq:reconrel1} and \eq{eq:reconrel2} hold for $O_b : \CH_{a} \to \CH_{a}$ and $O : \CH_{A} \to \CH_{A}$. The interpretation is clear; \emph{an operator $O_b$ that acts on $\CE[A]$ can be realised by an operator $O$ that acts on the boundary subregion $A$}. In other words, the bulk dual of boundary subregion is its entanglement wedge. Note that this constructive proof does not give an explicit procedure for reconstructing the entanglement wedge\cite{Dong:2016eik}, which is the problem this paper attempts to solve.

\subsection{Restriction to subregions of the boundary}
The Hilbert space of the whole boundary theory was given as span of states created from the vacuum by applying mode operators. Na{\"i}ve expectation for the boundary subregion Hilbert space is that it is given as the span of states created from the vacuum by applying boundary subregion mode operators; this is what happened for the bulk.

Unfortunately, it is in general not easy to find a procedure that can construct mode operators for the subregions at the boundary. Suppose that a boundary subregion $A$ of a time slice $\S$ is given. The subregion Hilbert space $\CH_A$ should describe physics on the boundary domain of dependence $\CD[A]$ of $A$. To construct the mode operators following the procedures of section \ref{sec:gffmodes}, construct a coordinate patch that covers $\CD[A]$ and Wick rotate the time coordinate to Euclidean signature. Then use a conformal transformation including Weyl rescaling that maps the ``past infinity'' of $\CD[A]$, defined as the most causally past point of $\CD[A]$, to the origin. The states in $\CH_A$ is then given by states generated by inserting operators at the origin via operator-state correspondence. In theory everything should work. In practice it is almost impossible to find such a map for a general subregion $A$.

On the other hand, it is possible to impose the following conditions on the truncated operators.
\bl
\tilde{\CO}(x) &= 0  && \text{for } x \in \CD [A^c]
\\ \la \tilde{\CO}(x_1) \tilde{\CO}(x_2) \cdots \tilde{\CO}(x_n) \ra &= \la \CO(x_1) \CO(x_2) \cdots \CO(x_n) \ra && \text{for } x_1, x_2, \cdots , x_n \in \CD [A]
\el
The first condition requires the truncated operators to be identically zero in the domain of dependence $\CD [A^c]$ of the complementary subregion $A^c = \S - A$. The second condition requires the truncated operators to be indistinguishable from the original operators in the domain of dependence $\CD [A]$ of the given subregion $A$. One possible way to satisfy both criteria is to write the truncated field $\tilde{\CO}(t,\O)$ of the boundary as the following mode expansion.
\bl
\tilde{\CO}(t,\O) &= \sum \CO_{nlm} \tilde{g}_{nlm}(t,\O) + \text{h.c} \label{eq:truncboundmodeexp}
\\ \tilde{g}_{nlm}(t,\O) &= \left\{
\begin{aligned}
&{g}_{nlm}(t,\O) && (t,\O) \in \CD[A]
\\ &0 && (t,\O) \in \CD[A^c]
\end{aligned} \right.
\el
$\CO_{nlm}$ and $g_{nlm}$ refers to the original boundary mode operators and mode functions. Note the similarity of this expansion to bulk mode expansion \eq{eq:truncmodeexp}. In view of this similarity, the mode function $\tilde{g}_{nlm}$ will be called the \emph{resummed mode function for the boundary}, although which resummation it originates from is totally vague. Since any analogue of wave equation does not exist at the boundary, there is no good criteria for determining the resummed boundary mode functions at boundary spacetime outside $\CD[A]$ and $\CD[A^c]$.

While the mode expansion \eq{eq:truncboundmodeexp} obscures the structure of the Hilbert space $\CH_A$ corresponding to the given boundary subregion $A$, it does capture the relevant physics in the given subregion. Therefore, the mode expansion \eq{eq:truncboundmodeexp} will be treated as a valid mode decomposition of the truncated operator $\tilde{\CO}$.

\subsection{Building smearing functions for the entanglement wedge} \label{sec:ewreconsubsec}
To construct the bulk, it remains to find resummed bulk mode functions $\tilde{f}_{nlm}$ that reduce to resummed boundary mode functions $\tilde{g}_{nlm}$ at the boundary by a slight variant of the last dictionary \eq{eq:dictionary3}.
\bl
\tilde{g}_{nlm}(t, \O) &\leftrightarrow \lim_{\r \to \pi/2} \left( \frac{\tilde{f}_{nlm}(t, \r, \O)}{\cos^\D \r} \right)  \label{eq:dictionary3var}
\el
As resummed boundary mode functions are well-defined only inside $\CD[A]$ and $\CD[A^c]$ of the boundary spacetime, it remains to find the bulk subregion $\S_A$ which gives the right boundary behaviour of the resummed bulk mode functions evaluated as $\fket{\tilde{f}_{nlm}} = \iden_A \fket{{f}_{nlm}}$. Define $\S$ as the time slice which includes causal surfaces\footnote{Causal surface $\xi_A$ of boundary subregion $A$ is defined as the ``rim'' of the causal wedge $\CW_C[A]$, or the intersection of the past and future bulk horizons of $\CD[A]$\cite{Almheiri:2014lwa}.} $\xi_A$ and $\xi_{A^c}$. Also define $\S_{cw}[A]$ as the intersection of $\S$ and the causal wedge $\CW_C[A]$ of $A$, i.e. $\S_{cw}[A] = \CW_C[A] \cap \S$. Then, any subregion $\S_{cw}[A] \subseteq \S_A \subseteq \S - \S_{cw}[A^c]$ satisfies the resummed mode function condition \eq{eq:dictionary3var} at $\CD[A]$ and $\CD[A^c]$. This ambiguity in choice of $\S_A$ is resolved by the fact that the bulk reconstructible from the given boundary subregions corresponds to the entanglement wedge $\CE[A]$ of boundary subregion $A$ \cite{Dong:2016eik}. This means the following requirement is imposed on the resummed bulk mode function $\tilde{f}_{nlm}$.
\bl
\tilde{f}_{nlm}(t,\r,\O) &= \left\{
\begin{aligned}
&{f}_{nlm}(t,\r,\O) && (t,r,\O) \in \CE[A]
\\ &0 && (t,\r,\O) \in \CE[A^c] 
\end{aligned} \right.
\el

Having determined the appropriate mode functions of the bulk, determining the smearing function can be proceeded in the manner described in section \ref{sec:bulkrecon}. Picking the region of boundary spacetime $\Xi$ to be equal to $\CD[A]$ in \ref{sec:boundprojdef}, the smearing function $K^A$ which reconstructs the bulk based on the boundary data in $\CD[A]$ can be written schematically as follows.
\bl
K^{A}(x|Y) &= \sum_{k} \fket{\tilde{f}_{k} (x)} P_{k, \CD[A]}^{+} (Y) + \fket{\tilde{f}_{k}^\ast (x)} P_{k, \CD[A]}^{-} (Y)
\el
The vectors $\fket{\tilde{f}_{k}}$ and $\fket{\tilde{f}_{k}^\ast}$ denote resummed mode functions of the bulk, and $P_{k, \CD[A]}^{+/-}$ denotes the integral kernel corresponding to the mode projection operator that projects out the positive/negative frequency mode operator that corresponds to the (set of) quantum number(s) $k$. Since details of this reconstruction are only a mere repetition of the procedure outlined in section \ref{sec:bulkrecon}, it will be omitted. Note that this process only requires existence of the dictionary between mode operators and resummed mode functions; if the dictionary can be built for other background spacetime, then it is possible to apply the same reconstruction method to that spacetime.

This construction sheds a new light onto the noninvertibility problem of the gram matrix $\bold{g}$ considered in section \ref{sec:feasibility}. If $\bold{g}$ is indeed invertible in general, it means that any state can be built solely from arbitrary subregions \emph{without} any cost of arbitrarily high precision\footnote{It is known that in QFT any state can be approximated by acting on the vacuum with local observables confined to a subregion of the total spacetime(Reeh-Schlieder property), which indicates presence of an enormous amount of entanglement\cite{Morrison:2014jha}.}, which seems to contradict the assumption that only the entanglement wedge is reconstructed; existence of $\bold{g}^{-1}$ means contribution of higher (in the sense of quantum numbers $n$ and $l$) modes to a lower mode projection operator diminishes sufficiently fast. This is another supporting evidence of the conclusion given in section \ref{sec:feasibility} that the construction should be viewed as an effective description rather than an exact one.

An interesting feature of this reconstruction mechanism is that when accessible boundary spacetime $\Xi = \CD[A]$ becomes smaller, lower modes become harder to discern. Since lower modes in the global patch of AdS are weighted to the centre of the AdS, this property is consistent with the expectation that a larger portion of the boundary is needed to probe deeper into the bulk. In \cite{Almheiri:2014lwa} this expectation was cast as the statement that operators of the deeper bulk are more robust against local erasures of the boundary. Another implication this observation poses is that in some cases disregarding the modes lower than a certain cut-off is needed for an efficient bulk reconstruction.

\subsection{Covariance under AdS isometries}
A relativistic system must obey covariance, but there is some subtlety regarding the construction proposed in this paper. It has been argued in section \ref{sec:feasibility} that non-invertibility of infinite dimensional matrices implies the need to introduce a cut-off. This cut-off is a UV cut-off at the boundary, which corresponds to an IR cut-off in the bulk. This fact raises doubts on covariance of the construction with respect to AdS isometries; the IR cut-off, the confining box of the system, will move around with the flow generated by the Killing vectors of AdS. Nevertheless, there are reasons to believe that covariance must remain. First of all, IR cut-off should not affect the local, microscopic details of the system and local causal structure is one of those details. Since relativity is about consistent description of causal structures, the subsystem corresponding to the entanglement wedge must be relativistic because the full system is relativistic. How is covariance under AdS isometries realised in this construction?

The smearing function obtained in \cite{Hamilton:2006az}, reproduced in \eq{eq:HKLLglobal}, has an explicit dependence on the radial coordinate $\r$ which is a frame-dependent quantity. This means the smearing function does not change as a scalar under AdS isometries.
\bl
K(x|Y) \to {K'}(x'|Y') &= K \left( x(x')|Y(Y') \right) \lim_{\r \to \pi/2} \left( \frac{\cos \r'}{\cos \r} \right)^{\D-d}
\\ &= K \left( x(x')|Y(Y') \right) \abs{\frac{\p Y'}{\p Y}}^{(\D/d)-1} \label{eq:smearingtrsf}
\el
The last line follows from the following identity which relates limiting values of cosines to Jacobian factors of boundary coordinates, which is proved in appendix \ref{app:Jacobian}.
\bl
\lim_{\r \to \pi/2} \left( \frac{\cos \r'}{\cos \r} \right)^{d} &= \abs{\frac{\p Y'}{\p Y}} \label{eq:JacobianIden}
\el

The covariant transformation rule \eq{eq:smearingtrsf} fits nicely with the fact that primary fields on the CFT(or GFT) side needs to be rescaled under coordinate transformations if two-point correlation function structure is required to be preserved in the new coordinates.
\bl
\CO_Y(Y) \to \CO_{Y'}(Y') &= \abs{\frac{\p Y}{\p Y'}}^{\D/d} \CO_{Y}\left( Y(Y') \right)
\el
The subscript under GFF $\CO$ is there to serve as a reminder that this field has the correct correlation function structure for the subscript coordinate system; $\CO_{Y}$ has the correct correlators when coordinate system $Y$ is used, and $\CO_{Y'}$ when $Y'$ is used. The transformation rule for the bulk scalar $\phi(x)$ is given by
\bl
\phi(x) \to \phi'(x') &= \phi \left( x(x') \right)
\el
and representation of the bulk scalar by boundary GFF is subject to the following transformation rule.
\bl
\phi'(x') &= \int {K'}(x'|Y') \CO_{Y'} dY'
\\ &= \phi \left( x(x') \right) = \int {K}\left( x(x') | Y \right) \CO_{Y} dY \nn
\\ &= \int {K}\left( x(x') | Y(Y') \right) \abs{\frac{\p Y'}{\p Y}}^{\D/d} \CO_{Y'} \abs{\frac{\p Y}{\p Y'}} dY' \nn
\\ &= \int {K}\left( x(x') | Y(Y') \right) \abs{\frac{\p Y'}{\p Y}}^{(\D/d)-1} \CO_{Y'} dY'
\el
Therefore imposing the transformation rule \eq{eq:smearingtrsf} to the smearing function constructed by the procedure proposed in this paper will guarantee covariance under AdS isometries, since integration over the new coordinates will be naturally recast into an integration over the coordinates that the smearing function was defined by. Note that in general the smearing function constructed in one reference frame would be different from the one constructed in another, as boundary mode functions $g_{k}$ and the gram matrix $g_{ij}$ constructed from them will be different in general. This is an explicit realisation of the ambiguities in projection operator $P^{m}$ mentioned in footnote \ref{fn:bdL2}.

\section{Discussion} \label{sec:discussion}
A procedure has been proposed in this paper which explicitly reconstructs a bulk scalar $\phi$ from its boundary dual operator $\CO$ when only a portion of the boundary data is given. The key idea is the observation that smearing function $K$ is an automation of reading mode operators at the boundary and assigning the corresponding bulk mode functions, so that it is possible to reconstruct the bulk whenever such a dictionary is given. The dictionary between mode operators and mode functions of the bulk and boundary for pure AdS perturbations was built to establish a working example for the construction. The bulk dual to subregions of the boundary was chosen to be the entanglement wedge, based on the recent proof\cite{Dong:2016eik} of bulk reconstruction. Resummed mode functions were used to obtain the mode functions for the entanglement wedge without explicitly solving the wave equation on a complicated coordinate patch that covers the entanglement wedge. Non-orthogonality of resummed mode functions was resolved by introducing a Gram matrix and its inverse, although an UV cut-off had to be introduced to resolve complications induced by infinite dimensionality of the Gram matrix. Because of the introduction of an UV cut-off, the smearing function constructed by this process is at best an effective reconstruction. In effect, the hard problem of solving the wave equation on a complicated coordinate patch was transferred to the problem of inverting a very large matrix, which is still hard but relatively tractable. Covariance under choice of mode functions guarantees that the construction is unique for a chosen reference frame.

There exist some subtleties that has not been adressed in the main article, one of which is the ambiguity of mode projection operators. As explained in footnote \ref{fn:bdL2}, there is some ambiguity in defining mode projection operators, which was essential for equivalence of smearing functions constructed in different reference frames. A criterion that seems to work that can kill this ambiguity is to require the $L^2$ norm of the mode projection operator, $\int_\Xi \abs{P^m}^2 dV$, to be minimised. On the other hand, it is rather obscure how this criterion can be imposed in practice; while it is possible to project out the orthogonal complement $V_1^\perp$ of vector space $V_1$ spanned by mode functions when some arbitrary cut-off is imposed, obtaining basis vectors of $V_1^\perp$ without introducing any cut-off is not readily available. Another subtlety that remains unresolved is consistent implementation of UV cut-off in the boundary and the bulk. The cut-off was introduced to regulate infinite dimensionality of the Gram matrix, so it is imposed in an ad hoc fashion. Had the cut-off been introduced in the beginning, there should be a consistent way of imposing the cut-off to the subregions so that dimensionality of the full Hilbert space factorise consistently. Although these subtleties do not seem to invalidate the procedure outlined in the main article, they do seem to provide some food for thought.

As the proposal to understand AdS/CFT as an error-correcting structure\cite{Almheiri:2014lwa} was one of the main motivations to study the problem of entanglement wedge reconstruction, it would be interesting to see how the structure manifests itself in the proposal of this paper. Reviewing the procedure of reconstruction, the only place that error-correcting structure can reside in seems to be the mode operator retrieval process. Note that there is nothing quantum inherent in this process; this is a classical signal retrieval process from data given by a non-orthogonal basis in absence of noise. A similar structure exists in the bulk theory; the mode expansion \eq{eq:truncmodeexp} shows that an arbitrary subregion of the bulk spacetime can encode mode operators of the global Fock states. This seems to suggest that such an error-correcting structure could be a generic feature of quantum field thoeries, which has its roots in Reeh-Schlieder property. In view of holographic code models\cite{Pastawski:2015qua,Yang:2015uoa}, this observation could be an indication that error correcting codes can also be built from non-hyperbolic tensor networks as well, a possibility already conjectured in \cite{Yang:2015uoa}.

Just like the original reconstructions of HKLL\cite{Hamilton:2005ju,Hamilton:2006az}, the bulk reconstruction suggested in this paper does not include $1/N$ corrections. Adding $1/N$ corrections implies interactions kick in, meaning that signal retrieval process described in section \ref{sec:boundprojdef} must include tolerance from errors induced by interactions. Whether such incorporation of errors can be used to understand the ``phase transition'' behaviour of the entanglement wedge\cite{Almheiri:2014lwa} would be an intersting problem to ponder on. Another facet of $1/N$ corrections is that non-perturbative objects such as solitons or instantons may appear at finite $N$. In presence of instantons, the total Hilbert space splits into a sum of different sectors. In such a case, it may not be possible to write the total Hilbert space as a product of subregion Hilbert spaces. While it was implicitly assumed that the total Hilbert space of the bulk and boundary can be decomposed into product of subregion Hilbert spaces, whether it is possible at all is a potential problem that this paper has avoided to answer. An indication of this problem already existed in \cite{Almheiri:2014lwa} when considering tripartitioning of the boundary to show that the bulk centre cannot be reconstructed in any of the subregions alone. This problem has also been raised in a somewhat different context in \cite{Donnelly:2016rvo}.

As a final remark, the causal structure of the entanglement wedge resembles the causal structure of long Einstein-Rosen bridges. It would be an interesting exercise to probe the implications of the reconstruction algorithm elaborated here to the problem of reconstructing the interior of long Einstein-Rosen bridges from boundary data. This problem will be left for future work.

\paragraph{Acknowledgements} The author would like to thank Daniel Harlow, Daniel Kabat, and Sangmin Lee for helpful discussions. This work was supported by the BK21 Plus Program(21A20131111123) funded by the Ministry of Education(MOE, Korea) and National Research Foundation of Korea(NRF).

\appendix
\section{Computation of boundary Jacobian factor} \label{app:Jacobian}
The goal of this appendix is to prove the identity \eq{eq:JacobianIden}.
\bl
\lim_{\r \to \pi/2} \left( \frac{\cos \r'}{\cos \r} \right)^{d} &= \abs{\frac{\p Y'}{\p Y}}
\el
This relation is based on the fact that bulk measure remains invariant under isometries of AdS$_{d+1}$.
\bl
dx = \sec^{d+1} \r~ d\r dY &\to dx' = \sec^{d+1} \r'~ d\r' dY' = dx
\el
$dY$ refers to the spacetime measure of the boundary; $dY = \sqrt{\abs{g_{\IR \times S^{d-1}}}} ~dt d\O$. Using Dirac delta and integrating over the radial coordinate $\r$ gives the following relation.
\bl
\int \delta \left[\sec{\r}(\r - \pi/2 + \e) \right] \sec^{d+1} \r~ d\r dY &= \left. \sec^{d} \r \right|_{\r= \pi/2 - \e}~ dY
\el
The factor of $\sec \r$ is included inside the argument of Dirac delta, since the \emph{proper length element} along the radial direction should be written as $\sec \r ~d\r$. The wanted identity is proved by combining the above relation with invariance of bulk measure and taking the limit $\e \to 0$.
\bl
\lim_{\r \to \pi/2} \sec^{d} \r' ~dY' &= \lim_{\r \to \pi/2} \sec^{d} \r ~dY
\\ dY' &= \lim_{\r \to \pi/2} \left( \frac{\cos \r'}{\cos \r} \right)^{d} ~dY = \abs{\frac{\p Y'}{\p Y}} dY
\el

As an example, consider AdS$_3$. Using the embedding space coordinates $u^\m$, AdS$_3$ of radius 1 can be given by the following equation.
\bl
-1 &= - \left( u^{-1} \right)^2 - \left( u^{0} \right)^2 + \left( u^{1} \right)^2 + \left( u^{2} \right)^2
\el
The following parametrisation of the hypersurface gives the global patch of AdS$_3$.
\bl
\begin{aligned}
u^{-1} &= \sec \r \cos t
\\ u^{0} &= \sec \r \sin t
\\ u^{1} &= \tan \r \cos \th
\\ u^{2} &= \tan \r \sin \th
\end{aligned} \label{eq:AdS3param}
\el
The global patch of AdS$_3$ has the following metric.
\bl
ds^2 = \sec^2 \r (-dt^2 + d\r^2 + \sin^2 \r ~d\th^2)
\el
Consider the following $u^{-1} - u^{1}$ plane rotation, which is an isometry of AdS$_3$. The primed coordinates are parametrised in the same way as in \eq{eq:AdS3param}.
\bl
\begin{aligned}
{u'}^{-1} &= \cosh \l ~u^{-1} + \sinh \l ~u^{1}
\\ {u'}^{1} &= \sinh \l ~u^{-1} + \cosh \l ~u^{1}
\end{aligned}
\el
After some manipulation, the following relation between primed and unprimed coordinates of AdS$_3$ can be obtained.
\bl
\begin{aligned}
\sec^2 \r' &= \sec^2 \r + 2 \cosh\l \sinh\l \sec\r \tan\r \cos\th \cos t 
\\ & \hskip 25pt + \sinh^2 \l (\sec^2 \r \cos^2 t + \tan^2 \r \cos^2 \th)
\\ \tan t' &= \frac{\sin t}{\cosh \l \cos t + \sinh \l \sin \r \cos \th}
\\ \tan \th' &= \frac{\sin \th}{\cosh \l \cos \th + \sinh \l \csc \r \cos t}
\end{aligned}
\el
The limit of the ratio of cosines can be directly evaluated by the first relation.
\bl
\lim_{\r \to \pi/2} \left( \frac{\cos \r}{\cos \r'} \right)^{2} = 1 + 2 \cosh\l \sinh\l \cos\th \cos t + \sinh^2\l (\cos^2 t + \cos^2 \th)
\el
The boundary is parametrised by $t, \th$ or $t', \th'$. The volume measure of the boundary is given by $dY = dt d\th$ and $dY' = dt' d\th'$. The relation between boundary coordinates can be deduced from the other two relations by taking the limit $\r \to \pi/2$.
\bl
\begin{aligned}
\tan t' &= \frac{\sin t}{\cosh \l \cos t + \sinh \l \cos \th}
\\ \tan \th' &= \frac{\sin \th}{\cosh \l \cos \th + \sinh \l \cos t}
\end{aligned}
\el
Direct evaluation of the Jacobian gives
\bl
\abs{\frac{\p(t',\th')}{\p(t,\th)}} = \frac{1}{1 + 2 \cosh\l \sinh\l \cos\th \cos t + \sinh^2\l (\cos^2 t + \cos^2 \th)}
\el
or
\bl
\lim_{\r \to \pi/2} \left( \frac{\cos \r'}{\cos \r} \right)^{2} &= \abs{\frac{\p(t',\th')}{\p(t,\th)}}
\el
which is consistent with \eq{eq:JacobianIden}.


\begin{thebibliography}{99}

\bibitem{Maldacena:1997re} 
  J.~M.~Maldacena,
  ``The Large N limit of superconformal field theories and supergravity,''
  Int.\ J.\ Theor.\ Phys.\  {\bf 38}, 1113 (1999)
  [Adv.\ Theor.\ Math.\ Phys.\  {\bf 2}, 231 (1998)]
  doi:10.1023/A:1026654312961
  [hep-th/9711200].

\bibitem{Hamilton:2005ju} 
  A.~Hamilton, D.~N.~Kabat, G.~Lifschytz and D.~A.~Lowe,
  ``Local bulk operators in AdS/CFT: A Boundary view of horizons and locality,''
  Phys.\ Rev.\ D {\bf 73}, 086003 (2006)
  doi:10.1103/PhysRevD.73.086003
  [hep-th/0506118].

\bibitem{Hamilton:2006az} 
  A.~Hamilton, D.~N.~Kabat, G.~Lifschytz and D.~A.~Lowe,
  ``Holographic representation of local bulk operators,''
  Phys.\ Rev.\ D {\bf 74}, 066009 (2006)
  doi:10.1103/PhysRevD.74.066009
  [hep-th/0606141].

\bibitem{Banks:1998dd} 
  T.~Banks, M.~R.~Douglas, G.~T.~Horowitz and E.~J.~Martinec,
  ``AdS dynamics from conformal field theory,''
  hep-th/9808016.

\bibitem{Bena:1999jv} 
  I.~Bena,
  ``On the construction of local fields in the bulk of AdS(5) and other spaces,''
  Phys.\ Rev.\ D {\bf 62}, 066007 (2000)
  doi:10.1103/PhysRevD.62.066007
  [hep-th/9905186].

\bibitem{Bousso:2012sj} 
  R.~Bousso, S.~Leichenauer and V.~Rosenhaus,
  ``Light-sheets and AdS/CFT,''
  Phys.\ Rev.\ D {\bf 86}, 046009 (2012)
  doi:10.1103/PhysRevD.86.046009
  [arXiv:1203.6619 [hep-th]].

\bibitem{Bousso:2012mh} 
  R.~Bousso, B.~Freivogel, S.~Leichenauer, V.~Rosenhaus and C.~Zukowski,
  ``Null Geodesics, Local CFT Operators and AdS/CFT for Subregions,''
  Phys.\ Rev.\ D {\bf 88}, 064057 (2013)
  doi:10.1103/PhysRevD.88.064057
  [arXiv:1209.4641 [hep-th]].

\bibitem{Parikh:2012kg} 
  M.~Parikh and P.~Samantray,
  ``Rindler-AdS/CFT,''
  arXiv:1211.7370 [hep-th].

\bibitem{Czech:2012be} 
  B.~Czech, J.~L.~Karczmarek, F.~Nogueira and M.~Van Raamsdonk,
  ``Rindler Quantum Gravity,''
  Class.\ Quant.\ Grav.\  {\bf 29}, 235025 (2012)
  doi:10.1088/0264-9381/29/23/235025
  [arXiv:1206.1323 [hep-th]].

\bibitem{Morrison:2014jha} 
  I.~A.~Morrison,
  ``Boundary-to-bulk maps for AdS causal wedges and the Reeh-Schlieder property in holography,''
  JHEP {\bf 1405}, 053 (2014)
  doi:10.1007/JHEP05(2014)053
  [arXiv:1403.3426 [hep-th]].

\bibitem{Almheiri:2014lwa} 
  A.~Almheiri, X.~Dong and D.~Harlow,
  ``Bulk Locality and Quantum Error Correction in AdS/CFT,''
  JHEP {\bf 1504}, 163 (2015)
  doi:10.1007/JHEP04(2015)163
  [arXiv:1411.7041 [hep-th]].

\bibitem{Mintun:2015qda} 
  E.~Mintun, J.~Polchinski and V.~Rosenhaus,
  ``Bulk-Boundary Duality, Gauge Invariance, and Quantum Error Corrections,''
  Phys.\ Rev.\ Lett.\  {\bf 115}, no. 15, 151601 (2015)
  doi:10.1103/PhysRevLett.115.151601
  [arXiv:1501.06577 [hep-th]].

\bibitem{Czech:2012bh} 
  B.~Czech, J.~L.~Karczmarek, F.~Nogueira and M.~Van Raamsdonk,
  ``The Gravity Dual of a Density Matrix,''
  Class.\ Quant.\ Grav.\  {\bf 29}, 155009 (2012)
  doi:10.1088/0264-9381/29/15/155009
  [arXiv:1204.1330 [hep-th]].

\bibitem{Dong:2016eik} 
  X.~Dong, D.~Harlow and A.~C.~Wall,
  ``Reconstruction of Bulk Operators within the Entanglement Wedge in Gauge-Gravity Duality,''
  Phys.\ Rev.\ Lett.\  {\bf 117}, no. 2, 021601 (2016)
  doi:10.1103/PhysRevLett.117.021601
  [arXiv:1601.05416 [hep-th]].

\bibitem{Bao:2016skw} 
  N.~Bao and I.~H.~Kim,
  ``Precursor problem and holographic mutual information,''
  arXiv:1601.07616 [hep-th].

\bibitem{Jafferis:2015del} 
  D.~L.~Jafferis, A.~Lewkowycz, J.~Maldacena and S.~J.~Suh,
  ``Relative entropy equals bulk relative entropy,''
  JHEP {\bf 1606}, 004 (2016)
  doi:10.1007/JHEP06(2016)004
  [arXiv:1512.06431 [hep-th]].

\bibitem{ElShowk:2011ag} 
  S.~El-Showk and K.~Papadodimas,
  ``Emergent Spacetime and Holographic CFTs,''
  JHEP {\bf 1210}, 106 (2012)
  doi:10.1007/JHEP10(2012)106
  [arXiv:1101.4163 [hep-th]].

\bibitem{Papadodimas:2012aq} 
  K.~Papadodimas and S.~Raju,
  ``An Infalling Observer in AdS/CFT,''
  JHEP {\bf 1310}, 212 (2013)
  doi:10.1007/JHEP10(2013)212
  [arXiv:1211.6767 [hep-th]].

\bibitem{Freivogel:2016zsb} 
  B.~Freivogel, R.~A.~Jefferson and L.~Kabir,
  JHEP {\bf 1604}, 119 (2016)
  doi:10.1007/JHEP04(2016)119
  [arXiv:1602.04811 [hep-th]].

\bibitem{Kabat:2011rz} 
  D.~Kabat, G.~Lifschytz and D.~A.~Lowe,
  ``Constructing local bulk observables in interacting AdS/CFT,''
  Phys.\ Rev.\ D {\bf 83}, 106009 (2011)
  doi:10.1103/PhysRevD.83.106009
  [arXiv:1102.2910 [hep-th]].

\bibitem{Kabat:2015swa} 
  D.~Kabat and G.~Lifschytz,
  ``Bulk equations of motion from CFT correlators,''
  JHEP {\bf 1509}, 059 (2015)
  doi:10.1007/JHEP09(2015)059
  [arXiv:1505.03755 [hep-th]].

\bibitem{Kabat:2016zzr} 
  D.~Kabat and G.~Lifschytz,
  ``Locality, bulk equations of motion and the conformal bootstrap,''
  JHEP {\bf 1610}, 091 (2016)
  doi:10.1007/JHEP10(2016)091
  [arXiv:1603.06800 [hep-th]].

\bibitem{Hamilton:2006fh} 
  A.~Hamilton, D.~N.~Kabat, G.~Lifschytz and D.~A.~Lowe,
  ``Local bulk operators in AdS/CFT: A Holographic description of the black hole interior,''
  Phys.\ Rev.\ D {\bf 75}, 106001 (2007)
  Erratum: [Phys.\ Rev.\ D {\bf 75}, 129902 (2007)]
  doi:10.1103/PhysRevD.75.106001, 10.1103/PhysRevD.75.129902
  [hep-th/0612053].

\bibitem{Leichenauer:2013kaa} 
  S.~Leichenauer and V.~Rosenhaus,
  ``AdS black holes, the bulk-boundary dictionary, and smearing functions,''
  Phys.\ Rev.\ D {\bf 88}, no. 2, 026003 (2013)
  doi:10.1103/PhysRevD.88.026003
  [arXiv:1304.6821 [hep-th]].

\bibitem{Rey:2014dpa} 
  S.~J.~Rey and V.~Rosenhaus,
  ``Scanning Tunneling Macroscopy, Black Holes, and AdS/CFT Bulk Locality,''
  JHEP {\bf 1407}, 050 (2014)
  doi:10.1007/JHEP07(2014)050
  [arXiv:1403.3943 [hep-th]].

\bibitem{Pastawski:2015qua} 
  F.~Pastawski, B.~Yoshida, D.~Harlow and J.~Preskill,
  ``Holographic quantum error-correcting codes: Toy models for the bulk/boundary correspondence,''
  JHEP {\bf 1506}, 149 (2015)
  doi:10.1007/JHEP06(2015)149
  [arXiv:1503.06237 [hep-th]].

\bibitem{Yang:2015uoa} 
  Z.~Yang, P.~Hayden and X.~L.~Qi,
  ``Bidirectional holographic codes and sub-AdS locality,''
  JHEP {\bf 1601}, 175 (2016)
  doi:10.1007/JHEP01(2016)175
  [arXiv:1510.03784 [hep-th]].

\bibitem{Donnelly:2016rvo} 
  W.~Donnelly and S.~B.~Giddings,
  ``Observables, gravitational dressing, and obstructions to locality and subsystems,''
  Phys.\ Rev.\ D {\bf 94}, no. 10, 104038 (2016)
  doi:10.1103/PhysRevD.94.104038
  [arXiv:1607.01025 [hep-th]].

\bibitem{Balasubramanian:2013lsa} 
  V.~Balasubramanian, B.~D.~Chowdhury, B.~Czech, J.~de Boer and M.~P.~Heller,
  ``Bulk curves from boundary data in holography,''
  Phys.\ Rev.\ D {\bf 89}, no. 8, 086004 (2014)
  doi:10.1103/PhysRevD.89.086004
  [arXiv:1310.4204 [hep-th]].

\bibitem{Headrick:2014eia} 
  M.~Headrick, R.~C.~Myers and J.~Wien,
  ``Holographic Holes and Differential Entropy,''
  JHEP {\bf 1410}, 149 (2014)
  doi:10.1007/JHEP10(2014)149
  [arXiv:1408.4770 [hep-th]].

\bibitem{Lin:2014hva} 
  J.~Lin, M.~Marcolli, H.~Ooguri and B.~Stoica,
  ``Locality of Gravitational Systems from Entanglement of Conformal Field Theories,''
  Phys.\ Rev.\ Lett.\  {\bf 114}, 221601 (2015)
  doi:10.1103/PhysRevLett.114.221601
  [arXiv:1412.1879 [hep-th]].

\bibitem{Nozaki:2013vta} 
  M.~Nozaki, T.~Numasawa, A.~Prudenziati and T.~Takayanagi,
  ``Dynamics of Entanglement Entropy from Einstein Equation,''
  Phys.\ Rev.\ D {\bf 88}, no. 2, 026012 (2013)
  doi:10.1103/PhysRevD.88.026012
  [arXiv:1304.7100 [hep-th]].

\bibitem{Lashkari:2013koa} 
  N.~Lashkari, M.~B.~McDermott and M.~Van Raamsdonk,
  ``Gravitational dynamics from entanglement 'thermodynamics',''
  JHEP {\bf 1404}, 195 (2014)
  doi:10.1007/JHEP04(2014)195
  [arXiv:1308.3716 [hep-th]].

\bibitem{Faulkner:2013ica} 
  T.~Faulkner, M.~Guica, T.~Hartman, R.~C.~Myers and M.~Van Raamsdonk,
  ``Gravitation from Entanglement in Holographic CFTs,''
  JHEP {\bf 1403}, 051 (2014)
  doi:10.1007/JHEP03(2014)051
  [arXiv:1312.7856 [hep-th]].

\bibitem{Swingle:2014uza} 
  B.~Swingle and M.~Van Raamsdonk,
  ``Universality of Gravity from Entanglement,''
  arXiv:1405.2933 [hep-th].

\bibitem{Lin:2015lfa} 
  J.~Lin,
  ``Bulk Locality from Entanglement in Gauge/Gravity Duality,''
  arXiv:1510.02367 [hep-th].

\bibitem{Maldacena:2015iua} 
  J.~Maldacena, D.~Simmons-Duffin and A.~Zhiboedov,
  ``Looking for a bulk point,''
  arXiv:1509.03612 [hep-th].

\bibitem{Czech:2016xec} 
  B.~Czech, L.~Lamprou, S.~McCandlish, B.~Mosk and J.~Sully,
  ``A Stereoscopic Look into the Bulk,''
  JHEP {\bf 1607}, 129 (2016)
  doi:10.1007/JHEP07(2016)129
  [arXiv:1604.03110 [hep-th]].

\bibitem{Engelhardt:2016wgb} 
  N.~Engelhardt and G.~T.~Horowitz,
  ``Towards a Reconstruction of General Bulk Metrics,''
  Class.\ Quant.\ Grav.\  {\bf 34}, no. 1, 015004 (2017)
  doi:10.1088/1361-6382/34/1/015004
  [arXiv:1605.01070 [hep-th]].

\bibitem{VanRaamsdonk:2010pw} 
  M.~Van Raamsdonk,
  ``Building up spacetime with quantum entanglement,''
  Gen.\ Rel.\ Grav.\  {\bf 42}, 2323 (2010)
  [Int.\ J.\ Mod.\ Phys.\ D {\bf 19}, 2429 (2010)]
  doi:10.1007/s10714-010-1034-0, 10.1142/S0218271810018529
  [arXiv:1005.3035 [hep-th]].

\bibitem{Kaplan:lec}
  J.~Kaplan, \textit{Lectures on AdS/CFT from the Bottom Up},
  \url{http://www.pha.jhu.edu/~jaredk/AdSCFTCourseNotesPublic.pdf}

\bibitem{Nastase:adscft}
  H.~N\u{a}stase, \textit{Introduction to the AdS/CFT Correspondence},
  Cambridge University Press (2015)

\bibitem{Diosi:info}
  L.~Di\'osi, \textit{A Short Course in Quantum Information Theory},
  Springer (2007)

\bibitem{Streater:pct}
  R.~F.~Streater and A.~S.~Wightman, \textit{PCT, Spin and Statistics, and All That},
  W. A. Benjamin, Inc. (1964)

\bibitem{Abramowitz:hmf}
  M.~Abramowitz and I.~A.~Stegun (Eds.), \textit{Handbook of Mathematical Functions: with Formulas, Graphs, and Mathematical Tables}, Tenth printing,
  National Bureau of Standards (1972)

\bibitem{NIST:DLMF}
  NIST Digital Library of Mathematical Functions,
  \url{http://dlmf.nist.gov/}

\bibitem{Richards:distr}
  J.~I.~Richards, H.~K.~Youn, \textit{Theory of Distributions: a non-technical introduction},
  Cambridge University Press (1990)

\end{thebibliography}
\end{document}